\documentclass[12pt]{article}
\usepackage{axodraw}
\usepackage{amssymb}
\usepackage{amsmath}
\usepackage{youngtab}
\usepackage{cite}
\usepackage[]{hyperref}
\input xy
\xyoption{all}

\numberwithin{equation}{section}

\begin{document}

\begin{center}

{\large\bf Notes on two-dimensional pure supersymmetric gauge theories}

\vspace{0.2in}

Wei Gu$^1$, Eric Sharpe$^2$, Hao Zou$^2$

\begin{tabular}{cc}
{\begin{tabular}{l}
$^1$ Center for Mathematical Sciences\\
Harvard University\\
Cambridge, MA  02138
\end{tabular}} &
{\begin{tabular}{l}
$^2$ Dep't of Physics\\
Virginia Tech\\
850 West Campus Dr.\\
Blacksburg, VA  24061
\end{tabular}}
\end{tabular}

{\tt weigu@cmsa.fas.harvard.edu},
{\tt ersharpe@vt.edu},
{\tt hzou@vt.edu}

\end{center}

In this note we 
study
IR limits of pure two-dimensional supersymmetric gauge theories with
semisimple
non-simply-connected gauge groups including $SU(k)/{\mathbb Z}_k$, 
$SO(2k)/{\mathbb Z}_2$, $Sp(2k)/{\mathbb Z}_2$, $E_6/{\mathbb Z}_3$,
and $E_7/{\mathbb Z}_2$ for various discrete
theta angles, both directly in the gauge theory and also in
nonabelian mirrors, extending a classification begun in previous work.  
We find in each case
that there are supersymmetric
vacua for precisely one value of the discrete theta angle, and no
supersymmetric vacua for other values,
hence supersymmetry is broken in the IR for most discrete theta angles.
Furthermore, for the one distinguished value of the discrete theta angle
for which supersymmetry is unbroken,
the theory has as many twisted chiral multiplet degrees of freedom 
in the IR 
as
the rank. 
We take this opportunity to further develop the technology of
nonabelian mirrors to discuss how the mirror to a $G$ gauge theory
differs from the mirror to a $G/K$ gauge theory for $K$ a subgroup of
the center of $G$.  In particular, the
discrete theta angles in these cases are considerably
more intricate than those of the pure gauge theories
studied
in previous papers, 
so we 
discuss
the realization of these more complex
discrete theta angles in the mirror construction.  
We find that discrete theta angles,
both in the original gauge theory and their mirrors,
are
intimately related to the
description of centers of universal covering groups as quotients of weight
lattices by root sublattices.  
We perform numerous consistency
checks, comparing results
against basic group-theoretic relations as well as with
decomposition, which describes how two-dimensional theories with
one-form symmetries (such as pure gauge theories with nontrivial centers)
decompose into disjoint unions, in this case of pure gauge theories with
quotiented gauge groups and discrete theta angles.

\begin{flushleft}
May 2020
\end{flushleft}

\newpage

\tableofcontents

\section{Introduction}

Two-dimensional pure (2,2) supersymmetric gauge theories with
gauge group $SU(k)$ were studied in \cite{Aharony:2016jki}, which argued
that the theory flows in the IR to a free theory of $k-1$ twisted chiral
multiplets.  In our previous work \cite{Gu:2018fpm,Chen:2018wep},
we looked at the nonabelian mirrors \cite{Gu:2018fpm} of these theories,
confirming the prediction of \cite{Aharony:2016jki}, and also extending to
other gauge theories.  Specifically, \cite{Gu:2018fpm} looked at two-dimensional
pure $SU(k)$, $SO(k)$, and $Sp(2k)$ gauge theories, and
\cite{Chen:2018wep} looked at pure $G_2$, $F_4$, and $E_{6,7,8}$ gauge
theories.  In each case, computations in the mirror suggested that the
mirror flows in the IR to a theory of twisted chiral multiplets,
as many as the rank of the theory, for precisely one value of the
discrete theta angle\footnote{
Briefly, a discrete theta angle weights nonperturbative sectors of the
path integral by phases 
determined by finite-group-valued degree-two characteristic classes, 
such as second Stiefel-Whitney classes $w_2$ in $(S)O(n)$ theories,
just as 
an ordinary theta angle weights nonperturbative sectors by phases
determined by instanton numbers (typically Chern and Pontryagin classes).
For semisimple gauge groups,
possible discrete theta angles in two-dimensional theories are determined
by the fundamental group of the gauge group.
See  
\cite{Hori:2011pd,Hori:1994uf,Hori:1994nc,Witten:1978ka,Gaiotto:2010be,Aharony:2013hda} 
for discussions of discrete theta angles in two-dimensional
theories.
} (in cases where the pure gauge theory has discrete
theta angles), and gave the distinguished value of the discrete
theta angle.

To be completely clear, the nonabelian mirror analysis of
\cite{Gu:2018fpm,Chen:2018wep} can explicitly determine for which
discrete theta angles supersymmetric vacua exist, and the number of
degrees of freedom remaining in the IR in those discrete theta angles.
However, the nonabelian mirror analysis of \cite{Gu:2018fpm,Chen:2018wep}
did not explicitly exclude
all possible interactions -- it can count the number of degrees of freedom,
and for the same reasons as in \cite{Aharony:2016jki} one may expect that
those degrees of freedom are free,
but the mirror construction 
does not explicitly demonstrate that the result is necessarily a free theory,
a limitation shared by the current work.

In any event, the analyses of \cite{Aharony:2016jki,Gu:2018fpm,Chen:2018wep}
still left a few groups unanalyzed,
namely further quotients
of theories above in cases in which the gauge group has a nontrivial
center.  These analyses require a deeper understanding of the
difference between the mirror to a $G$ gauge theory and a $G/K$ gauge
theory, for $K$ a subgroup of the center of $G$, and also an understanding
of the
mirrors of more intricate
discrete theta angles than described in \cite{Gu:2018fpm},
which is the reason they were not previously discussed.  In this paper,
we develop the technology of nonabelian mirrors further, to illustrate
how to distinguish the mirror to a $G$ gauge theory from the mirror of
a $G/$K gauge theory more generally, and by determining
mirrors to discrete theta angles in more complex cases.  By applying this
new understanding, we can analyze pure gauge theories with more general
gauge groups, and finish analyzing the cases that were not previously
studied in \cite{Aharony:2016jki,Gu:2018fpm,Chen:2018wep}.

We begin in section~\ref{sect:rev} with a number of general remarks.  We
give a general analysis of vacua of pure supersymmetric gauge theories
in the language of the original gauge theories, reducing questions about
discrete theta angles to some discrete choices of FI parameters, determined
up to log branch cut ambiguities.  In parallel, 
we also review the nonabelian mirror
proposal of \cite{Gu:2018fpm}, explain how to distinguish the mirror to
a $G$ gauge theory from that of a $G/K$ gauge theory,
and show how the gauge theory results can be derived
from the mirror.  To make the
discrete theta angle computations meaningful, in both the original
gauge theory and in the mirror, one needs invariants
on the set of discrete FI parameters modulo log branch cuts, which must
be computed on a case-by-case basis for each group.  In this section,
we describe in general terms how such computations of invariants can
be performed, and leave the details for each case for later sections.

In the remaining sections of this paper, we focus on two goals.
First, we apply nonabelian mirrors in detail in examples, to illustrate
the details of mirrors to gauge theories with gauge groups of the form
$G/K$ for $K$ a subgroup of the center of $G$, as this has not
been previously explained.  Second, for each 
case, we also compute the invariants needed to extract
human-readable discrete theta angles from sets of discrete choices of
FI parameters
modulo log branch cuts (corresponding to points on the weight lattice
modulo the root lattice).  In each case, we perform consistency
checks, comparing results for different representations of the
same group, and also comparing to the predictions of
decomposition, which we shall explain momentarily.

In section~\ref{sect:pure}, we discuss pure $SU(k)/{\mathbb Z}_k$
gauge theories.  As in all examples in this paper, we compute the single
discrete theta angle for which a pure $SU(k)/{\mathbb Z}_k$ gauge theory
has supersymmetric vacua, and give evidence that the theory flows in the
IR to a theory of as many twisted chiral multiplets as its rank.
For other discrete theta angles, the theory has no supersymmetric vacua,
and so supersymmetry is broken in the IR -- a pattern followed in all other
examples we study in this paper.

In section~\ref{sect:so2k} we discuss pure
$SO(2k)/{\mathbb Z}_2$ gauge theories.
Depending upon whether $k$ is even or odd, there is either
a ${\mathbb Z}_2 \times {\mathbb Z}_2$ or ${\mathbb Z}_4$ of discrete
theta angles.  We describe how these two possibilities arise in the mirror,
and compute for which discrete theta angles pure $SO(2k)/{\mathbb Z}_2$
theories have supersymmetric vacua, giving evidence that the theory flows
in the IR to a theory of as many twisted chiral multiplets as the rank.

In section~\ref{sect:sp2k} we turn to pure $Sp(2k)/{\mathbb Z}_2$ gauge
theories, repeating the same analyses as above.

In sections~\ref{sect:e6} and \ref{sect:e7} we turn to pure
$E_6/{\mathbb Z}_3$ and $E_7/{\mathbb Z}_2$ gauge theories.  For these
cases, to determine invariants computing discrete theta angles requires
more careful consideration of root and weight lattice structures, in a fashion
introduced in earlier sections.  Once those invariants are determined,
the rest of the analysis proceeds as before, determining for which 
discrete theta angles one has supersymmetric vacua, and checking that
results are consistent with decomposition.

One of the properties we check is `decomposition.'
This is a property of
two-dimensional gauge theories with finite subgroups that act
trivially on massless matter, or equivalently two-dimensional theories
with restrictions on nonperturbative sectors, have been discussed
in the literature since \cite{Pantev:2005rh,Pantev:2005zs}.
Such theories
possess discrete one-form symmetries, and so, as a result,
`decompose' into disjoint unions of simpler theories.
This was first discussed in \cite{Hellerman:2006zs}, and has since
been applied to understand exotic phases of certain abelian GLSMs
(see {\it e.g.}
\cite{Caldararu:2007tc,Hori:2011pd,Halverson:2013eua,Parsian:2018fhm}),
as well as to make predictions for
Gromov-Witten invariants of certain stacks, predictions which
were checked in {\it e.g.} \cite{ajt,tseng1,gt1}, see
\cite{Sharpe:2014tca} for further references.
See for example \cite{Sharpe:2019ddn} for a recent overview and
discussion in terms of one-form symmetries,
and \cite{Tanizaki:2019rbk} for an analogous decomposition in
four-dimensional theories with three-form symmetries.

The same ideas apply in principle to two-dimensional nonabelian
gauge theories in which the center of the gauge group acts trivially
on massless matter, as discussed in \cite{Sharpe:2014tca}. 
The `nonabelian decomposition' discussed there says that a two-dimensional
$G$ gauge
theory with center $Z(G)$ and center-invariant massless matter
decomposes into a disjoint union of 
$G/Z(G)$ gauge theories with various discrete theta angles.
The simplest version of this is, schematically, the statement
\begin{displaymath}
SU(2) \: = \: SO(3)_+ + SO(3)_-,
\end{displaymath}
where $SU(2)$ and $SO(3)$ indicate gauge theories and the subscripts
indicate discrete theta angles.  This decomposition and its
analogues have been checked in many examples of many different kinds
of two-dimensional theories with one-form symmetries, and we will
encounter many examples in this paper, in which we will sue decomposition
as a consistency check on existence of supersymmetric vacua for
various discrete theta angles.    

Finally, we mention that other proposals for nonabelian mirrors
have appeared in the math community in
e.g.
\cite{rietsch1,teleman}, as reviewed in \cite{Gu:2018fpm}[section 4.9,
appendix A].

\section{Review and overview}
\label{sect:rev}

In this section we will discuss vacua of pure gauge theories for
general (connected) gauge groups, 
both in the original gauge theory and the mirror.
We will give general results in both cases, illustrating how all
supersymmetric vacua arise in a theory with a unique discrete theta angle,
and that the IR limit consists of as many twisted chiral multiplet degrees
of freedom as the rank of the gauge group.  (For other discrete theta
angles, there are no supersymmetric vacua, hence supersymmetry is broken in the
IR.)  However, to actually determine
the discrete theta angle for which supersymmetric vacua exist
in a more meaningful, human-readable fashion in each case
will require further group-dependent work, which will be detailed
in later sections.

\subsection{Gauge theory}

Consider a two-dimensional (2,2) supersymmetric gauge theory
with gauge group $G$, where we for the moment assume $G$ is semisimple,
connected and simply-connected.

For this theory, on the Coulomb branch, there is an effective 
one-loop twisted superpotential given by
\cite[equ'ns (2.17), (2.19)]{Nekrasov:2014xaa},
\cite[equ'n (4.77)]{Closset:2015rna},
\cite[equ'n (2.37)]{Closset:2017vvl}
\begin{eqnarray}
W_{\rm eff} & = &
- \sum_a \sum_i \sigma_a \rho^a_i \ln\left(  \sum_b
\sigma_b \rho^b_i \: - \: \tilde{m}_i\right) \: + \: 
\sum_a \sum_i \sigma_a \rho^a_i
\: - \: \sum_a \sigma_a t_a
\nonumber \\
& & 
\: - \: \sum_a \sum_{\tilde{\mu}} \sigma_a \alpha^a_{\tilde{\mu}}
\left[ \ln \left( \sum_b \sigma_b \alpha^b_{\tilde{\mu}} \right)
\: - \: 1 \right]
\nonumber \\
& & 
\: + \: \sum_i \tilde{m}_i \ln \left(
\sum_b
\sigma_b \rho^b_i \: - \: \tilde{m}_i\right) ,
\end{eqnarray}
where $\sigma_a$ are the adjoint-valued scalar of the vector multiplet,
in a maximal torus, $\rho^a_i$ are the weights of the matter representations
appearing in the theory, $\alpha^a_{\tilde{\mu}}$ are the roots of the
Lie algebra ${\mathfrak g}$ of the gauge group $G$, and
$\tilde{m}_i$ are twisted masses.  For $G$ semisimple, the $t_a$ encode
discrete theta angles, as we shall describe momentarily.

In principle, one should also take a Weyl-group orbifold of the
effective theory of the $\sigma_a$, and exclude certain points on the
space of $\sigma$'s (the ``excluded locus'').  For a connected gauge group,
the Weyl orbifold group acts freely on the allowed $\sigma$, and so we
shall suppress those details in this analysis.

The terms involving the root vectors are often simplified as
\begin{eqnarray}
\lefteqn{
\sum_{\tilde{\mu}=1}^{n-r} \alpha_{\tilde{\mu}}^a 
\left( \ln \left( \sum_b \sigma_b \alpha^b_{\tilde{\mu}} \right) \: - \: 1
\right)
} \nonumber \\
& = &
\sum_{\rm pos'} \alpha_{\tilde{\mu}}^a 
\ln \left( \sum_b \sigma_b \alpha^b_{\tilde{\mu}} \right) 
\: - \:
\sum_{\rm pos'} \alpha_{\tilde{\mu}}^a \left( \ln \left( \sum_b \sigma_b \alpha^b_{\tilde{\mu}} \right) - \pi i \right),
\\
& = &
\sum_{\rm pos'} i \pi  \alpha_{\tilde{\mu}}^a,
\label{eq:genl-corr:simp1}
\end{eqnarray}
giving a shift of the theta angle matching that given in
\cite{Hori:2013ika}[equ'n (10.9)].
This expression glosses over branch cut subtleties arising in
non-simply-connected semisimple groups, as we shall see shortly.

Now, suppose $K$ is a subgroup of the center of $G$,
and we have $K$-invariant matter.  We can then consider a $G/K$-gauge
theory with the same matter, for which we can work out an analogue
of the expression above.  For simplicity, as in this paper we focus
on pure gauge theories, let us consider the case that there is no matter,
and hence no twisted masses.  The effective one-loop superpotential in
this case is
\begin{equation}
W_{\rm eff} \: = \:
 - \: \frac{1}{|K|} \sum_a \sum_{\tilde{\mu}} \sigma_a \alpha^a_{\tilde{\mu}}
\left[ \ln \left( \sum_b \sigma_b \alpha^b_{\tilde{\mu}} \right)
\: - \: 1 \right]
\: - \: \sum_a \sigma_a t_a,
\label{eq:gauge:pure-w}
\end{equation}
where the factor of $1/|K|$ arises from the fact that a maximal torus
of $G/K$ is $|K|$ times smaller than that of $G$.

In principle, because of the logarithms in the 
expression~(\ref{eq:gauge:pure-w}) above,
there are branch cuts, which have the effect of shifting the $t_a$.
Specifically, across a branch cut in the logarithm associated to
$\alpha^a_{\nu}$, we see
\begin{equation}
t_a \: \mapsto \: t_a \: + \: \frac{2 \pi i}{|K|} \alpha^a_{\nu}.
\end{equation}
Thus, we see that the $t_a$ are only defined up to shifts by
root lattice vectors (rescaled by $2 \pi i/|K|$),
and in particular are not uniquely defined.

Now, let us consider how to describe the discrete theta angles.
Just as an ordinary theta angle weights nonperturbative sectors by phases,
a discrete theta angle also weights nonperturbative sectors by phases.
The difference is that discrete theta angles couple to 
a characteristic class in $H^2(\Sigma, \pi_1(G))$ (over
two-dimensional spacetime $\Sigma$).  Specifically,
a discrete theta angle is a character of $\pi_1(G)$, giving a $U(1)$
phase for every value of the characteristic class above.
Physically, along the Coulomb branch, we are restricting to bundles
whose structure groups are reducible to a maximal torus $T \subset G$.
For such bundles,
the characteristic class in $H^2(\Sigma, \pi_1(G))$ can be
represented by a collection of first Chern classes $c_1$
(as many as the dimension of $T$, the rank of $G$),
and so we can represent the discrete theta angles by a collection of
ordinary theta angles, as many as the rank of $G$,
encoded in the $t_a$.
Mathematically, as we shall see later in 
equation~(\ref{eq:disc-theta:root-weight}),
there is a short exact sequence relating
\begin{equation}
1 \: \longrightarrow \:
\mbox{root lattice} \: \longrightarrow \: \mbox{weight lattice}
\: \longrightarrow \: \mbox{characters} \: \longrightarrow \: 1
\end{equation}
We can therefore lift the discrete theta angles from characters to
weight lattice elements, which correlate with the ordinary theta
angles encoded in $t_a$ along the Coulomb branch.

To summarize,
for a given discrete theta angle, we take the $t_a$ to be
(proportional to)
elements of the weight lattice, up to shifts by the root lattice.
Later in subsection~\ref{subsect:discrete-theta}
we will elaborate on how 
this will enable us to identify the $t_a$ with discrete theta angles.

Setting aside the precise meaning of the $t_a$ for a moment, we now
turn to finding the vacua in the gauge theory.  Using the
simplification~(\ref{eq:genl-corr:simp1}), it is straightforward
to read off from~(\ref{eq:gauge:pure-w}) the equations of motion for $\sigma_a$,
which are simply a constraint on the $t_a$:
\begin{equation}  \label{eq:gauge:t-crit}
t_a \: = \: - \frac{\pi i}{|K|} \sum_{\rm pos'}  \alpha_{\tilde{\mu}}^a,
\end{equation}
with no constraint on the $\sigma_a$.  Thus, for precisely
the value of the discrete
theta angle determined by the $t_a$ above (modulo log branch cuts),
the pure $G/K$ gauge theory has supersymmetric vacua,
described by as many twisted chiral multiplet degrees of freedom as the
number of $\sigma$ fields.  (To go further and determine whether this
is a free theory would require knowledge of the metric, which we do not
have.)

In passing, in the original pure $G$ gauge theory, one has a very
similar result.  The effective superpotential is nearly the same
as~(\ref{eq:gauge:pure-w}), albeit without the factor of $1/|K|$.
The equations of motion imply
\begin{equation}
t_a \: = \: - \pi i \sum_{\rm pos'}  \alpha_{\tilde{\mu}}^a,
\end{equation}
with no constraint on the $\sigma_a$.
Here, the $t_a$ are weight lattice elements times $2 \pi i$,
and are only defined up to root lattice shifts
\begin{equation}
t_a \: \mapsto \: t_a \: + \: 2 \pi i \alpha^a_{\nu}
\end{equation}
for various $\nu$.  If $G$ is not simply-connected, then there will
be discrete theta angles to be found amongst the possible $t_a$,
in the same fashion as we will describe shortly.

To finish the analysis, we need to understand more precisely how to map
the $t_a$ to discrete theta angles.  In the next several sections of this
paper, we will compute the invariants on the weight lattice quotients
that compute discrete theta angles from values of $t_a$, to give 
a more meaningful answer.  Before doing that analysis, we will now turn
to the mirror theories, to see how the corresponding analysis is performed
there.  We will see that the mirror analysis is very similar to the analysis
in the original gauge theory.

\subsection{Mirror}

Given a two-dimensional (2,2) supersymmetric $G$-gauge theory,
$G$ connected, with
matter chiral superfields in a (typically reducible) representation $\rho$,
it was proposed in \cite{Gu:2018fpm} that the mirror is a 
Weyl-group orbifold of a Landau-Ginzburg
model with fields
\begin{itemize}
\item $\sigma_a$, $a \in \{1, \cdots, r \equiv {\rm rank}\, G \}$,
\item $Y_i$, $i \in \{1, \cdots, N \equiv {\rm dim}\, \rho\}$,
each of periodicity $2 \pi i$ as in \cite[section 3.1]{Hori:2000kt},
\item $X_{\tilde{\mu}}$, corresponding to nonzero roots of the Lie algebra
${\mathfrak g}$ of $G$,
\end{itemize}
and superpotential
\begin{eqnarray}
W & = & \sum_{a=1}^r \sigma_a \left( 
\sum_{i=1}^N \rho_i^a Y_i \: - \: \sum_{\tilde{\mu}=1}^{n-r} \alpha_{\tilde{\mu}}^a 
\ln X_{\tilde{\mu}}
\: - \: t_a \right) 
\nonumber \\
& & \: - \:
\sum_{i=1}^N  \tilde{m}_i 
\left( Y_i \: - \: \sum_a \rho_i^a t_a \right) 
\nonumber \\
& & \: + \: \sum_{i=1}^N \exp\left(-Y_i\right) \: + \:
\sum_{\tilde{\mu}=1}^{n-r} X_{\tilde{\mu}}, 
\label{eq:proposal-w}
\end{eqnarray}
where $n$ is the dimension of $G$,
the $\rho_i^a$ are the weight vectors for the representation
$\rho$, and the $\alpha_{\tilde{\mu}}^a$ are root vectors for the
Lie algebra ${\mathfrak g}$ of $G$.  The $\tilde{m}_i$ are
twisted masses.  The $t_a$ are mirror to Fayet-Iliopoulos
parameters and/or discrete theta angles, depending upon $G$.
(We also sometimes write
$Z = - \ln X$ for notational convenience.)  This proposal satisifes a wide
variety of tests, including reproducing the quantum cohomology/Coulomb
branch relations, excluded loci, and dualities of two-dimensional
nonabelian gauge theories studied in \cite{Hori:2011pd} and correlation
functions obtained from supersymmetric localization 
(see e.g. \cite{Closset:2017vvl}), and has
been extended to (0,2) supersymmetric theories \cite{Gu:2019byn}. 
For connected gauge groups, the Weyl group orbifold acts freely on the
critical locus, so as we only consider connected gauge groups in this paper,
we shall largely ignore the orbifold, except to take
Weyl-invariant combinations of fields.
(Mirrors to gauge theories whose gauge groups have multiple components were
discussed in \cite{Gu:2019zkw,Gu:2020nub}.)

Let us elaborate on the mirrors of the Fayet-Iliopoulos parameters, the $t_a$.
In cases in which the gauge group has a $U(1)$ factor, the corresponding
linear combination of the $t_a$'s decouples from the $\ln X$ branch
cut ambiguities, and is a free parameter.  In other cases,
the $t_a$ are constrained by those same $\ln X$ branch cuts, meaning that
in crossing a branch cut, the $(t_a)$ are shifted by $2 \pi i$ times a root
lattice vector.  Specifically, under
\begin{equation}
X_{\tilde{\mu}} \: \mapsto \: X_{\tilde{\mu}} \exp(2 \pi i ),
\end{equation}
the $t_a$ are shifted by
\begin{equation}
t_a \: \mapsto \: t_a \: - \: 2 \pi i  \alpha^a_{\tilde{\mu}}.
\end{equation}
Only those linear combinations of the $t_a$ which can be
`disentangled' from the $\alpha$'s can be free parameters; the rest are
only defined up to addition of root lattice vectors.  
We take the discrete theta angles to be defined by $(t_a)$ living on\footnote{
In principle, from our description, each $t_a$ lives on a torus,
which we interpret as the mirror to a theta angle in the $U(1)^r$
gauge theory on the Coulomb branch.  In this paper we only consider
discrete theta angles, hence we restrict to $t_a$ on the weight lattice
(mod shifts by root lattice vectors).
}
$2 \pi i$ times
the weight lattice, modulo shifts by the root lattice.

The mirror to a $G/K$ gauge theory of the same form,
with matter assumed $K$-invariant, is given by
a Landau-Ginzburg model with the same fields and the superpotential
\begin{eqnarray}
W & = & \frac{1}{|K|} \sum_{a=1}^r \sigma_a \left( 
\sum_{i=1}^N \rho_i^a Y_i \: - \: \sum_{\tilde{\mu}=1}^{n-r} \alpha_{\tilde{\mu}}^a 
\ln X_{\tilde{\mu}}
\: - \: t_a \right) 
\nonumber \\ 
& & \: - \: 
\sum_{i=1}^N  \tilde{m}_i 
\left( Y_i \: - \: \sum_a \rho_i^a t_a \right) 
\nonumber \\ 
& & \: + \: \sum_{i=1}^N \exp\left(-Y_i\right) \: + \:
\sum_{\tilde{\mu}=1}^{n-r} X_{\tilde{\mu}}, 
\label{eq:proposal-w-modk}
\end{eqnarray}
where for semisimple $G$ the $(t_a)$ are now taken to live on
$2 \pi i/|K|$ times the weight lattice, and log branch cuts generate
shifts by root lattice elements in the form
\begin{equation}
t_a \: \mapsto \: t_a \: - \: \frac{2 \pi i}{|K|} \alpha^a_{\tilde{\mu}}.
\end{equation}
In other words, to describe the mirror to a $G/K$ gauge theory,
where $G$ is semisimple and simply-connected,
we begin with the mirror to
the $G$ gauge theory, divide the weight and root lattice vectors
by a factor of $|K|$, and then the discrete theta angles are encoded
in $t_a \in (2 \pi i / |K|) {\cal W}$, modulo shifts by elements of
$(2 \pi i/|K|) {\cal R}$, where ${\cal W}$ denotes the weight lattice
and ${\cal R}\subset {\cal W}$ the root lattice.

If we were to add matter that is not invariant under $K$,
and set aside for the moment the fact that the original gauge theory
would not be well-defined,
then due to the $Y$ periodicities, the mirror's parameters $t_a$ would
be shifted, and so the discrete theta angles would not be well-defined.
Only if there is no matter that transforms under $K$ are $G/K$
discrete theta angles defined, as indeed the $G/K$ gauge theory itself
is not well-defined in that case.

Now, let us turn to a formal analysis of the vacua in the mirror to a
pure $G/K$-gauge theory.
The mirror superpotential has the form
\begin{equation}
W \: = \: - \frac{1}{|K|} \sum_a \sigma_a \sum_{\tilde{\mu}}
\alpha^a_{\tilde{\mu}} \ln X_{\tilde{\mu}} \: - \:
\sum_a \sigma_a t_a \: + \: \sum_{\tilde{\mu}} X_{\tilde{\mu}}.
\end{equation}
The critical locus equations
\begin{equation}
\frac{\partial W}{\partial X_{\tilde{\mu}}} \: = \: 0 \: = \:
\frac{\partial W}{\partial \sigma_a}
\end{equation}
imply
\begin{eqnarray}
X_{\tilde{\mu}} & = & - \frac{1}{|K|} \sum_a \sigma_a \alpha^a_{\tilde{\mu}},
\\
t_a & = & - \frac{1}{|K|} \sum_{\tilde{\mu}} \alpha^a_{\tilde{\mu}}
\ln X_{\tilde{\mu}}.
\end{eqnarray}
It is straightforward to show that
\begin{equation}
\sum_{\tilde{\mu}} \alpha^a_{\tilde{\mu}} \ln X_{\tilde{\mu}}
\: = \: i \pi \sum_{\rm pos'} \alpha^a_{\tilde{\mu}},
\end{equation}
hence we find on the critical locus that
\begin{equation}
t_a \: = \: - \frac{\pi i}{|K|} \sum_{\rm pos'} \alpha^a_{\mu},
\end{equation}
which matches equation~(\ref{eq:gauge:t-crit}) that was derived from the
pure gauge theory.  As there, these $t_a$ are only defined up to log branch
cuts, so to extract a human-readable expression for the discrete theta
angle, one must compute lattice invariants, as we shall do in examples
later in this paper.

The remainder of the analysis now follows the same form as in
\cite{Gu:2018fpm}:  along the critical locus, there are as many
twisted chiral degrees of freedom as the rank of $G/K$, and the
superpotential vanishes.  As in the gauge theory, our analysis comes
slightly short of necessarily demonstrating that the IR limit is a free
field theory, as for example we do not have any control over the 
metric.  However, our results are certainly consistent with getting
a set of free fields, for the one value of the discrete theta angle
determined by the $t_a$ above.  (For other discrete theta angles,
the $G/K$ gauge theory has no supersymmetric vacua, as is visible
both in the mirror as well as in the previous analysis of the original
gauge theory.  Thus, for most discrete theta angles, we see that the
pure gauge theory breaks supersymmetry in the IR.)

\subsection{Discrete theta angles}
\label{subsect:discrete-theta}

So far we have given a general analysis of vacua in pure
$G/K$ gauge theories and their mirrors, and come to the same conclusion:
for precisely one value of the discrete theta angle, there are
supersymmetric vacua (in fact, a family), and no supersymmetric
vacua (hence supersymmetry breaking)
for other discrete theta angles.  However, our expressions for the discrete
theta angles, in terms of the $t_a$, are subject to log branch cut
ambiguities, and so by themselves are not very illuminating.  To
better understand how to interpret the $t_a$, we need to construct
invariants, maps from the weight lattice modulo the root lattice to the
finite groups describing the discrete theta angles more usefully.
In this section, we will describe the general procedure for constructing
such maps, and in later sections of this paper, we will construct them 
explicitly and perform consistency checks on the results.

For semisimple groups, we have described the $t_a$ as (proportional to)
weight lattice elements, modulo shifts by root lattice elements generated
by log branch cuts.
Mathematically, for semisimple Lie groups,
the quotient of weight and root lattices corresponds\footnote{
This is a standard old result.  See for example
\cite[section V.7]{btd} for an overview.
} to
the center of the universal cover of the gauge group.  In other words,
for any Lie group $G$, if $\tilde{G}$ denotes its universal cover,
then the center $Z(\tilde{G})$ 
coincides with the weight lattice modulo shifts by
root sublattice vectors.
Then, the fundamental group of $\tilde{G}/Z(\tilde{G})$ is 
$Z(\tilde{G})$, and corresponds 
\cite{Hori:2011pd,Hori:1994uf,Hori:1994nc,Witten:1978ka,Gaiotto:2010be,Aharony:2013hda} 
to possible discrete theta
angles in a two-dimensional $\tilde{G}/Z(\tilde{G})$ gauge theory.
It is for this reason that
we can associate the $t_a$ with discrete theta
angles.

More formally, if we let $\tilde{T}$ denote a maximal torus in
$\tilde{G}$ and $T$ its projection into $\tilde{G}/Z(\tilde{G})$,
we have a commuting diagram
\begin{equation}
\xymatrix{
1 \ar[r]  & Z( \tilde{G} ) \ar[r] 
\ar@{=}[d] & \tilde{G} \ar[r] & \tilde{G}/Z(\tilde{G}) \ar[r] & 1
\\
1 \ar[r] & Z(\tilde{G}) \ar[r] & \tilde{T} \ar@{^{(}->}[u] \ar[r] &
T \ar@{^{(}->}[u] \ar[r] & 1,
}
\end{equation}
hence
\begin{equation}
\xymatrix{
1 \ar[r] & {\rm Hom}(T, U(1)) \ar[r] &
{\rm Hom}(\tilde{T},U(1)) \ar[r] &
{\rm Hom}( Z(\tilde{G}), U(1)) \ar[r] & 1.
} \label{eq:disc-theta:root-weight}
\end{equation}
In passing, Hom$(T,U(1))$ can be identified with the root lattice and
Hom$(\tilde{T},U(1))$ with the weight lattice.
Here again we see the possible discrete theta angles as defining the
difference between the weight spaces of $\tilde{G}$ and 
$\tilde{G}/Z(\tilde{G})$.  We also see that in principle the weights of
one are a rescaling of the weights of the other by $| Z(\tilde{G}) |$,
which we will also encounter in examples.

Concretely, for semisimple gauge factors,
if we let $A$ denote an integer matrix encoding a basis of root vectors in the
weight vector space, then using integer row reduction, we can diagonalize
it, with the diagonal entries determining the center and the basis determining
invariants.  Specifically, there exist
invertible integer matrices $P$, $Q$ (necessarily of determinant $\pm 1$)
such that
\begin{equation}
Q^{-1} A P \: = \: D,
\end{equation}
where $D$ is a diagonal integer matrix, chosen so that the diagonal
entries are $\{ d_1, \cdots, d_r \}$, with all $d_i \in 
{\mathbb Z}$, $d_i \geq 1$, and $d_i \leq d_{i+1}$. 
(Technically, 
this is known as a Smith decomposition
of $A$.)
Then,
the center is ${\mathbb Z}_{d_1} \times {\mathbb Z}_{d_2} \times
\cdots \times {\mathbb Z}_{d_r}$. 
Furthermore, one can construct a map
to any ${\mathbb Z}_d$ factor as follows.  Let $v = [x_1, \cdots, x_r]^T$
denote the basis element corresponding to diagonal entry $d$, 
then in the basis in which
$A$ is diagonal, we can map weight vectors to ${\mathbb Z}_d$ by 
contracting a weight vector with the corresponding eigenvector, and
taking the result mod $d$.  We shall see this in examples later.
In this fashion, we can construct invariants which map
the $t_a$ to elements of a finite group, and so read off the
discrete theta angle in a useful form.

In the rest of this paper, we will explore the realizations of
discrete theta angles via constructing such invariants,
and complete our
analyses of two-dimensional pure (2,2)-supersymmetric gauge theories.

\section{Pure $SU(k)/{\mathbb Z}_k$ gauge theories}
\label{sect:pure}

In \cite{Aharony:2016jki}, it was suggested that a two\footnote{
In passing, for four-dimensional analogues, the reader may find
\cite{Aharony:2013hda,Ang:2019txy} interesting.
} dimensional
(2,2) supersymmetric pure $SU(k)$ gauge theory should flow to a theory
of $k-1$ free twisted
chiral multiplets.  This was verified from nonabelian
mirrors in \cite[section 12.3]{Gu:2018fpm}.
In this section, we will use nonabelian mirrors to analyze the
IR behavior of pure $SU(k)/{\mathbb Z}_k$ gauge theories, and see that
supersymmetry is unbroken in the IR for precisely one value of the
discrete theta angle.
We have already given a quick general analysis of vacua, but we still need to 
compute lattice invariants needed to give a human-readable version of the
discrete theta angles, and at the same time, we will also walk through
the mirror computations in greater detail, as they may be less familiar
to many readers.
Furthermore, we will check the
result by applying nonabelian decomposition
\cite{Hellerman:2006zs,Sharpe:2014tca,Sharpe:2019ddn}.  Pure $SU(k)$ gauge
theory decomposes into a disjoint union of $SU(k)/{\mathbb Z}_k$ gauge
theories with different theta angles, and we shall see that $k-1$
multiplets arise in exactly one of those $SU(k)/{\mathbb Z}_k$ theories,
whereas component theories with other discrete theta angles have
no supersymmetric vacua at all.

Before describing the general case, we will first review the
special cases of $SU(2)$ and $SU(3)$ theories.

\subsection{$SU(2)/{\mathbb Z}_2 = SO(3)$ theory}
\label{sect:su2}

The case of pure $SU(2)$ and $SO(3)$ theories has already been
discussed in \cite[section 12.1]{Gu:2018fpm}, but we will quickly
review the mirror to $SO(3)$ cases as a warm-up exercise.

It is convenient to
describe this in terms of
two $\tilde{\sigma}$ fields obeying the constraint
\begin{equation}  \label{eq:sum-sigma}
\sum_a \tilde{\sigma}_a \: = \: 0.
\end{equation}
The mirror superpotential to a pure $SO(3)$ theory then takes the form
\cite{Gu:2018fpm}
\begin{equation}
W \: = \:
(1/2) \sum_a \tilde{\sigma}_a \sum_{\mu \neq \nu} \alpha^a_{\mu \nu} Z_{\mu \nu}
\: + \: t \tilde{\sigma}_1
\: + \: \sum_{\mu \neq \nu} X_{\mu \nu}, 
\end{equation}
where
\begin{displaymath}
\alpha^a_{\mu \nu} \: = \: - \delta^a_{\mu} + \delta^a_{\nu},
\end{displaymath}
and where $t \in \{0, \pi i\}$ encodes the discrete theta angle in the
theory.  This can be simplified to
\begin{equation}
W \: = \: \tilde{\sigma}_1 \left( \ln X_{12} - \ln X_{21} \right) \: + \:
t \tilde{\sigma_1} \: + \: X_{12} \: + \: X_{21}.
\end{equation}
The critical loci are given by
\begin{eqnarray}
\frac{ X_{12} }{ X_{21} } & = & \exp(-t),
\\
X_{12} & = & - 2 \tilde{\sigma}_1 \: = \: - X_{21}.
\end{eqnarray}
In the case $t = 0$, these equations contradict one another, so there is
no solution, and no supersymmetric vacuum.  In the case $t = \pi i$,
these equations consistently require $X_{12} = - X_{21}$.
After integrating out $\sigma_1$, the superpotential becomes $W = 0$,
and identifying $X_{12}$ with $-X_{21}$, we see that we have one field
remaining.  

Now, to be clear, to completely verify the claim of 
\cite{Aharony:2016jki} in this case, more is required.
We have explicitly computed for which discrete theta angle supersymmetric
vacua can exist, and we have verified that for that one discrete
theta angle, there is one field remaining, a twisted chiral multiplet degree
of freedom, for which the superpotential vanishes.  
However, reference \cite{Aharony:2016jki} claimed that the IR theory is
a free twisted chiral multiplet, and although we can explicitly
check that there is one twisted chiral multiplet degree of freedom,
with vanishing superpotential, we do not have enough control over
the K\"ahler potential to exclude the possibility of a nontrivial
kinetic term.  Thus, although we can confirm that for the correct
discrete theta angle the expected number of degrees of freedom are present,
we cannot explicitly exclude the possibility of a nontrivial interaction.
Our results are therefore certainly consistent with the claim of
\cite{Aharony:2016jki}, and we can confirm many details, but not all.

\begin{figure}[ht]
\centering
\begin{picture}(350,20)(0,0)
\Line(0,10)(350,10)
\CCirc(75,10){3}{Black}{}
\CCirc(175,10){3}{Black}{}
\CCirc(275,10){3}{Black}{}
\CArc(125,10)(3,0,360)
\CArc(225,10)(3,0,360)
\CArc(25,10)(3,0,360)
\CArc(325,10)(3,0,360)
\end{picture}
\caption{Schematic illustration of weight and root lattice for $SU(2)$.
Solid circles indicate points on both lattices; empty circles indicate
points on the weight lattice that are not also on the root lattice.
\label{fig:su2}
}
\end{figure}
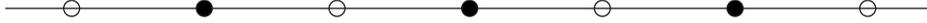

Now let us recast this description in terms of 
the general formalism described in
section~\ref{sect:rev}, to illustrate concretely how the abstract definition
works in this case.  First, the weight lattice corresponds to the
integers ${\mathbb Z}$, and the root lattice to the even integers
$2 {\mathbb Z}$.  This is implicit in the mirror construction.
For example, for a single fundamental,
the mirror superpotential
contains the terms
\begin{equation}
\sum_a \tilde{\sigma}_a \rho^a_b Y_b
\: = \: 
\tilde{\sigma}_1 \left( Y_1 - Y_2 \right),
\end{equation}
using the identity~(\ref{eq:sum-sigma}).
Thus, fields correspond to weights as follows:
\begin{equation}
Y_1 \sim +1, \: \: \:
Y_2 \sim -1,
\end{equation}
and the lattice generated by these weights is precisely the integers.
Analogously, from the roots, the superpotential of the
mirror to an $SU(2)$ gauge theory
contains the terms
\begin{equation}
\sum_a \tilde{\sigma}_a \alpha^a_{\mu \nu} Z_{\mu \nu}
\: = \: \tilde{\sigma}_1 \left( \alpha^1_{\mu \nu} - \alpha^2_{\mu \nu} \right)
Z_{\mu \nu}
\: = \: \tilde{\sigma}_2 \left( -2 Z_{12} + 2 Z_{21} \right),
\end{equation}
from which we see that fields correspond to weights as follows:
\begin{equation}
Z_{21} \sim +2, \: \: \:
Z_{12} \sim -2,
\end{equation}
generating $2 {\mathbb Z} \subset {\mathbb Z}$.
We have sketched this one-dimensional lattice in
figure~\ref{fig:su2}.

To get the $SO(3)$ mirror, we divide the weights by a factor of
$| Z(SU(2)) | = | {\mathbb Z}_2 |$, and then the possible discrete theta 
angles emerge as choices $t \in 2 \pi i {\mathbb Z}/2$ 
modulo root lattice shifts by elements of $2 \pi i (2/2) {\mathbb Z}$.
That leaves the possible values of $t$ as $\{ 0, \pi i \}$, modulo
shifts by $2 \pi i$, agreeing with the description reviewed above from 
\cite{Gu:2018fpm}.

For completeness, in terms of the general expression~(\ref{eq:gauge:t-crit})
for the values of $t_a$ for which the theory has supersymmetric vacua,
plugging in $|K| = 2$ and $\alpha = 2$, we have
\begin{equation}
t \: = \: - \frac{\pi i}{2} \left( 2 \right) \: = \: - \pi i,
\end{equation}
which given the periodicity 
\begin{equation}
t \: \sim \: t \: + \: \frac{2 \pi i }{2} \left( 2 \right) \: = \:
t \: + \: 2 \pi i,
\end{equation}
corresponds to the nontrivial discrete theta angle, matching the
computation above.

To summarize, we have reviewed how a pure supersymmetric $SO(3)$
gauge theory in two dimensions has supersymmetric vacua for one value
of the discrete theta angle, but not the other, and how there is one 
associated degree of freedom in the IR.

Now, let us briefly compare to the prediction of
decomposition 
\cite{Pantev:2005rh,Pantev:2005zs,Hellerman:2006zs,Sharpe:2014tca,Sharpe:2019ddn}.
Briefly, this predicts that a pure two-dimensional supersymmetric
$SU(2)$ gauge theory will decompose as a disjoint union of two $SO(3)$
theories with each discrete theta angle, or schematically
\begin{equation}
SU(2) \: = \: SO(3)_+ \: + \: SO(3)_-.
\end{equation}
In \cite{Aharony:2016jki,Gu:2018fpm} it was argued that the pure $SU(2)$
gauge theory will flow in the IR to a theory of one twisted chiral
multiplet, and using mirrors, we have reviewed how precisely one of the
two $SO(3)$ theories will do the same, while the other has no supersymmetric
vacua at all.  Thus, our results are consistent
with decomposition.

In passing, we should add that these results are consistent with
elliptic genus computations in \cite[appendix A]{Kim:2017zis}.
There, it was shown that the elliptic genus of both the pure $SU(2)$
and the $SO(3)_-$ theories have the same elliptic genus, namely
\begin{equation}
\frac{ \theta_1( \tau | -z ) }{ \theta_1( \tau | -2z) },
\end{equation}
while the elliptic genus of $SO(3)_+$ is identically zero.
Thus, if we let $Z$ denote elliptic genera, we have
\begin{equation}
Z( SU(2) ) \: = \: Z( SO(3)_+ ) \: + \: Z( SO(3)_- ),
\end{equation}
as expected from decomposition,
and furthermore, since the elliptic genus of $SO(3)_+$ vanishes,
we have immediately that the Witten index of this theory vanishes,
which is consistent with the computation of supersymmetry breaking
in $SO(3)_+$ above.

\subsection{$SU(3)/{\mathbb Z}_3$ theory}
\label{sect:su3}

Now, we will repeat the same analysis for the pure $SU(3)/{\mathbb Z}_3$ theory,
which has a few technical complications relative to $SU(2)$ and $SO(3)$.

Following section~\ref{sect:rev}, we begin with the mirror to a pure
$SU(3)$ theory, which was given in \cite[section 12.2]{Gu:2018fpm} 
as a Landau-Ginzburg model with superpotential
\begin{eqnarray}
W & = &
\tilde{\sigma}_1 \left( - Z_{12} - Z_{13} + Z_{21} + Z_{31} \right)
\nonumber \\
& & \: + \:
\tilde{\sigma}_2 \left( - Z_{21} - Z_{23} + Z_{12} + Z_{32} \right)
\nonumber \\
& & \: + \:
\tilde{\sigma}_3 \left( - Z_{31} - Z_{32} + Z_{13} + Z_{23} \right)
\nonumber \\
& & \: + \:
X_{12} + X_{13} + X_{21} + X_{23} + X_{31} + X_{32},
\\ \nonumber \\
& = &
\tilde{\sigma}_1 \left( - Z_{12} + Z_{21} - 2 Z_{13} + 2 Z_{31} +
Z_{32} -  Z_{23} \right)
\nonumber \\
& & \: + \:
\tilde{\sigma}_2 \left( - Z_{21} + Z_{12} - 2 Z_{23} + 2 Z_{32} +
Z_{31} - Z_{13} \right)
\nonumber \\
& & \: + \:
X_{12} + X_{13} + X_{21} + X_{23} + X_{31} + X_{32},
\end{eqnarray}
where we have used the constraint
\begin{equation}
\sum_a \tilde{\sigma}_a \: = \: 0
\end{equation}
to write
\begin{equation}
\tilde{\sigma}_3 \: = \: - \tilde{\sigma}_1 - \tilde{\sigma}_2.
\end{equation}
In these conventions, the weight lattice of $SU(3)$ is ${\cal W} = 
{\mathbb Z}^2$,
which we can see as follows.   The mirror to a single fundamental
contains the superpotential terms
\begin{equation}
\sum_a \tilde{\sigma}_a \rho^a_b Y_b
\: = \:
\tilde{\sigma}_1 \left( Y_1 - Y_3 \right) \: + \:
\tilde{\sigma}_2 \left( Y_2 - Y_3 \right),
\end{equation}
(using equation~(\ref{eq:sum-sigma}),)
from which we deduce the following correspondence between fields and
weights on the weight lattice:
\begin{equation}
Y_1 \sim (1,0), \: \: \:
Y_2 \sim (0,1), \: \: \:
Y_3 \sim (-1,-1).
\end{equation}
These span the lattice ${\mathbb Z}^2$. 
Similarly, the superpotential terms corresponding to the roots 
gives the following correspondence between fields and weights:
\begin{equation}
Z_{21} \sim (1,-1), \: \: \:
Z_{31} \sim (2,1), \: \: \:
Z_{32} \sim (1,2),
\end{equation}
\begin{equation}
Z_{12} \sim (-1,1), \: \: \:
Z_{13} \sim (-2,-1), \: \: \:
Z_{23} \sim (-1,-2).
\end{equation}
As expected, roots span a sublattice of the weight lattice.
Points on the root sublattice ${\cal R}$ include, for example,
\begin{eqnarray}
(0,3) & = & (1,2) + (-1,1),
\\
(3,0) & = & (2,1) + (1,-1),
\end{eqnarray}
and points which are not on the root sublattice include
\begin{equation}
(1,0), (2,0), (1,1), \: \: \:
(0,1), (0,2), (-1,-1).
\end{equation}

Following the general analysis of section~\ref{sect:rev},
the mirror to the pure $SU(3)/{\mathbb Z}_3$ theory should be obtained by
multiplying the lattice vectors by a factor of $1/| {\mathbb Z}_3 | = 1/3$,
and also adding potential discrete theta angles, which correspond to
a set of Fayet-Iliopoulos-like parameters $(t_a) \in (2 \pi i/3) {\cal W}$
modulo shifts by elements of $(2 \pi i/3) {\cal R}$.
Concretely, write each
\begin{equation}
t_a \: = \: m_a \frac{2 \pi i}{3}
\end{equation}
for some integer $m_a$.
The individual values of $m_a$ and $t_a$ are not uniquely defined -- in addition
to root lattice shifts,
the Weyl group can rotate individual choices into one another,
without changing the physics.
The overall quantity 
\begin{equation}  \label{eq:su3-invt}
\sum_a m_a \mod 3,
\end{equation}
however, is both permutation (Weyl)-invariant
and also invariant under root lattice shifts (which shift the sum by
multiples of $3$).  We can therefore take this sum mod 3 to be a compact
representation of the discrete theta angle.

We can recover this invariant more formally using the procedure
discussed in section~\ref{sect:rev}.  One matrix that encodes
a basis of root vectors in the weight lattice is
\begin{equation}
A \: = \: \left[ \begin{array}{cc}
2 & 1 \\
1 & 2 \end{array} \right].
\end{equation}
It is straightforward to check that 
\begin{equation}
\left[ \begin{array}{rr} 1 & -1 \\ -1 & 2 \end{array} \right]
\left[ \begin{array}{cc} 2 & 1 \\ 1 & 2 \end{array} \right]
\left[ \begin{array}{cc} 1 & 1 \\ 0 & 1 \end{array} \right]
\: = \:
\left[ \begin{array}{cc} 1 & 0 \\ 0 & 3 \end{array} \right].
\end{equation}
This is consistent with
the result that $SU(3)$ has center ${\mathbb Z}_3$, and the map 
from the weight lattice to ${\mathbb Z}_3$ is given by contracting with
the basis element corresponding to the last diagonal entry:
\begin{equation}
\left( m_a \right) \: \mapsto \: (1,1) \cdot (m_1, m_2) \mod 3
\: = \: m_1 + m_2 \mod 3,
\end{equation}
agreeing with the result above.

Putting this together,
the mirror to a pure (supersymmetric) $SU(3)/{\mathbb Z}_3$ 
gauge theory with any choice of discrete
theta angle
is defined by the superpotential
\begin{eqnarray}
W & = &
(1/3) \tilde{\sigma}_1 \left( - Z_{12} - Z_{13} + Z_{21} + Z_{31} \right)
\nonumber \\
& & \: + \:
(1/3) \tilde{\sigma}_2 \left( - Z_{21} - Z_{23} + Z_{12} + Z_{32} \right)
\nonumber \\
& & \: + \:
(1/3) \tilde{\sigma}_3 \left( - Z_{31} - Z_{32} + Z_{13} + Z_{23} \right)
\nonumber \\
& & \: + \: t_1 \tilde{\sigma}_1 \: + \: t_2 \tilde{\sigma}_2 \: + \:
X_{12} + X_{13} + X_{21} + X_{23} + X_{31} + X_{32},
\\ \\
& = &
(1/3) \tilde{\sigma}_1 \left( - Z_{12} + Z_{21} - 2 Z_{13} + 2 Z_{31} +
Z_{32} -  Z_{23} \right)
\nonumber \\
& & \: + \:
(1/3) \tilde{\sigma}_2 \left( - Z_{21} + Z_{12} - 2 Z_{23} + 2 Z_{32} +
Z_{31} - Z_{13} \right)
\nonumber \\
& & \: + \: t_1 \tilde{\sigma}_1 \: + \: t_2 \tilde{\sigma}_2 \: + \:
X_{12} + X_{13} + X_{21} + X_{23} + X_{31} + X_{32},
\end{eqnarray}
where
\begin{equation}
t_1, t_2 \: \in \: \{ 0, 2 \pi i/3, 4 \pi i/3 \}
\end{equation}
define a discrete theta angle via the invariant~(\ref{eq:su3-invt}), 
and where we have for convenience
written the result in terms of three $\tilde{\sigma}$'s, related by
\begin{equation}
\sum_a \tilde{\sigma}_a \: = \: 0.
\end{equation}
The critical locus is given by
\begin{equation}  \label{eq:su3:antisymm}
X_{12} \: = \: - X_{21}, \: \: \:
X_{13} \: = \: - X_{31}, \: \: \:
X_{23} \: = \: - X_{32},
\end{equation}
\begin{equation}
X_{13} \: - \: X_{23} \: = \: X_{12}.
\end{equation}
\begin{eqnarray}
\left[  \left( \frac{ X_{12} }{X_{21}} \right)
\left( \frac{ X_{13} }{ X_{31} } \right)^2
\left( \frac{ X_{23} }{ X_{32} } \right)
\right]^{1/3} & = & \exp(-t_1),
\label{eq:su3:t1} \\
\left[ \left( \frac{ X_{21} }{ X_{12} } \right)
\left( \frac{ X_{23} }{ X_{32} } \right)^2
\left( \frac{ X_{13} }{ X_{31} } \right)
\right]^{1/3} & = & \exp(-t_2).
\label{eq:su3:t2}
\end{eqnarray}

Let us check that the branch cuts in the last pair of equations are
compatible with the definition of the discrete theta angles given earlier.
From the critical locus equations, we know that $X_{12}/X_{21} = -1$.
If we write $X_{12} / X_{21} = \exp(+ i \pi)$, then change to
$\exp(-i \pi)$, then $(X_{12}/X_{21})^{1/3}$ changes by 
$\exp(- 2 \pi i/3)$, hence $m_1$ increases by $1$.  At the same time,
$m_2$ decreases by $1$, for the same reason, hence we see that the sum over
$m_i$ is invariant under choices of branch cuts.  If we perform the
same transformation on $X_{13}/X_{31}$, $m_1$ increases by $2$ and
$m_2$ increases by $1$ -- so the sum of the $m_i$ is invariant mod $3$.
The same change on $X_{23}/X_{32}$ increases $m_1$ by $1$ and $m_2$ by $2$,
so that again the sum of the $m_i$ is invariant mod $3$.  Thus,
the discrete theta angle (as defined above) does not depend upon choices
of branch cuts.

Picking the branch $X_{ij}/X_{ji} = \exp(i \pi)$ for $i<j$, we see that 
equations~(\ref{eq:su3:antisymm}), (\ref{eq:su3:t1}),
(\ref{eq:su3:t2}) have a solution when
\begin{equation}
\exp(-t_1) \: = \: \exp(4 \pi i/3), \: \: \:
\exp(-t_2) \: = \: \exp(2 \pi i/3),
\end{equation}
which corresponds to the case $m_1 = -2, m_2 = -1$, hence
\begin{equation}
m_1 + m_2 \: \equiv \: 0 \mod 3,
\end{equation}
so in order for the critical locus to be nonempty,
the discrete theta angle must vanish.

More compactly, we can use the invariant~(\ref{eq:su3-invt}) and
compute on the critical locus
\begin{equation}
\exp(- t_1 - t_2) \: = \:
\frac{ X_{13} }{ X_{31} } \frac{ X_{23} }{ X_{32} } \: = \: +1,
\end{equation}
which manifestly does not have any root-branch-cuts, and demonstrates
explicitly that one only has supersymmetric vacua for vanishing discrete
theta angle in this case.

For that one choice of discrete theta angle, the critical locus is
nonempty.  Integrating out the $X$ fields as in
\cite[section 12]{Gu:2018fpm} reduces the theory to a set of $k-1$ 
$\tilde{\sigma}$ fields (of which one takes Weyl-orbifold-invariant
combinations), with vanishing superpotential --
two (plausibly IR free) 
twisted chirals in the IR.
For other choices of discrete theta angle, the equations for the critical
locus have no solutions, and there is no supersymmetric vacuum,
hence supersymmetry is broken in the IR.

For completeness, let us also verify that the same result can be
obtained from equation~(\ref{eq:gauge:t-crit}).  Taking the positive
roots to be
\begin{equation}
(1,-1), \: \: \:
(2,1), \: \: \:
(1,2), 
\end{equation}
we have
\begin{equation}
t_1 \: = \: - i \pi \frac{4}{3}, \: \: \:
t_2 \: = \: - i \pi \frac{2}{3},
\end{equation}
hence
\begin{equation}
t_1 + t_2 \: = \: - 2 \pi i \: \sim \: 0,
\end{equation}
matching the computation above, that supersymmetric vacua exist in this
case only for the trivial ${\mathbb Z}_3$ discrete theta angle.

Now, let us apply nonabelian decomposition 
\cite{Hellerman:2006zs,Sharpe:2014tca,Sharpe:2019ddn}.  We write, schematically,
\begin{equation}
SU(3) \: = \: \left( \frac{ SU(3) }{ {\mathbb Z}_3 } \right)_0
\: + \: \left( \frac{ SU(3) }{ {\mathbb Z}_3 } \right)_1
\: + \: \left( \frac{ SU(3) }{ {\mathbb Z}_3 } \right)_2,
\end{equation}
where the subscripts indicate discrete theta angles (written
as integers mod 3).
We argued in \cite[section 12.2]{Gu:2018fpm} that a pure $SU(3)$
gauge theory seems to flow in the IR to two twisted chiral superfields,
and we have argued in this section that for precisely one value of
the discrete theta angle (the trivial case), a pure $SU(3)/{\mathbb Z}_3$
gauge theory also flows in the IR to two twisted chiral superfields.
Pure $SU(3)/{\mathbb Z}_3$ gauge theories for nontrivial discrete theta 
angles have no supersymmetric vacua (so supersymmetry is broken in the
IR).  Thus, we find our results are
consistent with decomposition:  the IR behavior of a pure $SU(3)$
theory matches that of the disjoint union above of pure
$SU(3)/{\mathbb Z}_3$ gauge theories.

\subsection{General $SU(k)/{\mathbb Z}_k$ theories}
\label{sect:suk}

Now, let us turn to mirrors of pure $SU(k)/{\mathbb Z}_k$ theories 
for more general $k$.  Following the general analysis of 
section~\ref{sect:rev}, we start with the mirror to an $SU(k)$ gauge
theory (given in \cite[sections 8, 12]{Gu:2018fpm}), rescale the lattices
by a factor of $1/|{\mathbb Z}_k| = 1/k$, and add $t_a$ to allow for the
possibility of discrete theta angles.

From \cite[sections 8, 12]{Gu:2018fpm},
the mirror to a pure $SU(k)$ gauge theory is a Landau-Ginzburg model
with superpotential
\begin{eqnarray}
W & = &
 \tilde{\sigma}_1 \bigl[  - Z_{12} - Z_{13} - \cdots - Z_{1 (k-1)}
- 2 Z_{1k} - Z_{2k} - \cdots - Z_{(k-1)k}
\nonumber \\
& & \hspace*{0.75in}
+ Z_{21} + Z_{31} + \cdots + Z_{(k-1)1} + 2 Z_{k1} +
Z_{k2} + \cdots + Z_{k(k-1)} \bigr]
\nonumber \\
& & \: + \:
 \tilde{\sigma}_2 \bigl[ - Z_{21} - Z_{23} - Z_{24} - \cdots -
Z_{2(k-1)} - 2 Z_{2k} - Z_{1k} - Z_{3k} - \cdots - Z_{(k-1)k}
\nonumber \\
& & \hspace*{0.75in}
+ Z_{12} + Z_{32} + Z_{42} + \cdots + Z_{(k-1) 2} + 2 Z_{k2}
+ Z_{k1} + Z_{k3} + Z_{k4} + \cdots + Z_{k(k-1)} \bigr]
\nonumber \\
& & \: + \: \cdots
 \: + \: \sum_{\mu \neq \nu} X_{\mu \nu}.
\end{eqnarray}
In these conventions, the weight lattice of $SU(k)$ is
${\cal W} = {\mathbb Z}^{k-1}$.  For example, the mirror superpotential
to a single fundamental of $SU(k)$ has terms
\begin{equation}
\sum_a \tilde{\sigma}_a \rho^a_b Y_b \: = \: 
\tilde{\sigma}_1 Y_1 + \tilde{\sigma}_2 Y_2 + \cdots +
\tilde{\sigma}_{k-1} Y_{k-1} + \left( - \tilde{\sigma}_1 - \tilde{\sigma}_2
- \cdots - \tilde{\sigma}_{k-1} \right) Y_k,
\end{equation}
from which we deduce the following correspondence between fields and
weights on the root lattice:
\begin{equation}
Y_1 \sim (1,0, \cdots, 0), \: \: \:
Y_2 \sim (0,1,0, \cdots, 0), \: \: \: \cdots, \: \: \:
Y_k \sim (-1, \cdots, -1).
\end{equation}
These span ${\cal W} = {\mathbb Z}^{k-1}$.
Following the same procedure,
we associate $Z_{\mu \nu}$ and roots as follows:
\begin{eqnarray}
Z_{12} & \sim & (-1, +1, 0, \cdots, 0),  \label{eq:suk:z12}
\\
Z_{13} & \sim & (-1, 0, +1, 0, \cdots, 0),
\\
\cdots & & \nonumber \\
Z_{1, k-1} & \sim & (-1, 0, \cdots, 0, 1),
\\
Z_{1k} & \sim & (-2, -1, \cdots, -1),  \label{eq:suk:z1k}
\end{eqnarray}
and so forth for others.  This defines a sublattice of the weight lattice,
as expected.  Points on the root lattice include
\begin{equation}
(k,0,\cdots,0) \: \sim \: Z_{21} + Z_{31} + \cdots + Z_{k-1.1} + Z_{k1},
\end{equation}
but not
\begin{equation}
(1,0,\cdots,0), \: \: \: 
(2,0,\cdots,0), \: \: \: \cdots, \: \: \:
(k-1,0,\cdots,0),
\end{equation}
for example.
More generally, points on the root sublattice ${\cal R}$ have the
distinguishing property that the sum of their coordinates (in the basis above)
is an integer multiple of $k$.

Following the general analysis of section~\ref{sect:rev}, the mirror to
a pure $SU(k)/{\mathbb Z}_k$ theory should be obtained by multiplying the
lattice vectors by a factor of $1/| {\mathbb Z}_k | = 1/k$, and also adding
potential discrete theta angles, which correspond to a set of 
Fayet-Iliopoulos-like parameters $(t_a) \in (2 \pi i/k) {\cal W}$ modulo
shifts by elements of $(2 \pi i/k) {\cal R}$.  Concretely, write each
\begin{equation}
t_a \: = \: m_a \frac{2 \pi i}{k}
\end{equation}
for some integer $m_a$.  The individual values of $m_a$ and $t_a$ are not
uniquely defined -- in addition to root lattice shifts, the Weyl group can
rotate individual choices into one another, without changing
the physics.  However, the quantity
\begin{equation}
\sum_a m_a \mod k
\end{equation}
is invariant under both permutations (the Weyl group) as well
as under root lattice shifts (which add multiples of $k$ to the sum).
We can use this invariant to compactly compute the discrete theta angle
corresponding to a given set of $t_a$, and so we take the discrete
theta angle to be
\begin{equation}
n \mod k \: \equiv \: \sum_a m_a \mod k.
\end{equation}

We can also derive this invariant using the general procedure
discussed in section~\ref{sect:rev}.  Consider, for example,
the following $(k-1) \times (k-1)$ matrix of root lattice basis vectors:
\begin{equation}
\left[ \begin{array}{ccccc}
2 & 1 & 1 & \cdots & 1 \\
1 & 2 & 1 & \cdots & 1 \\
1 & 1 & 2 & \cdots & 1 \\
\vdots & & & \dots & \\
1 & 1 & 1 & \cdots & 2 \end{array} \right].
\end{equation}
We can integer diagonalize as follows:
\begin{eqnarray}
\lefteqn{
\left[ \begin{array}{rrrrr}
1 & 0 & 0 & \cdots & -1 \\
0 & 1 & 0 & \cdots & -1 \\
0 & 0 & 1 & \cdots & -1 \\
\vdots & & & \cdots &  \\
-1 & -1 & -1 & \cdots & k
\end{array} \right]
\left[ \begin{array}{ccccc}
2 & 1 & 1 & \cdots & 1 \\
1 & 2 & 1 & \cdots & 1 \\
1 & 1 & 2 & \cdots & 1 \\
\vdots & & & \dots & \\
1 & 1 & 1 & \cdots & 2 \end{array} \right]
\left[ \begin{array}{ccccc}
1 & 0 & 0 & \cdots & 1 \\
0 & 1 & 0 & \cdots & 1 \\
0 & 0 & 1 & \cdots & 1 \\
\vdots & & & \cdots &  \\
0 & 0 & 0 & \cdots & 1
\end{array} \right]
}
\nonumber \\
& & \hspace*{2in}
\: = \:
\left[ \begin{array}{ccccc}
1 & 0 & 0 & \cdots & 0 \\
0 & 1 & 0 & \cdots & 0 \\
0 & 0 & 1 & \cdots & 0 \\
\vdots & & & \cdots &  \\
0 & 0 & 0 & \cdots & k
\end{array} \right],
\end{eqnarray}
naturally generalizing the $SU(3)$ discussion.
Almost all of the diagonals of this matrix are $1$; the one
different diagonal is $k$.  The basis element corresponding to diagonal $k$
is
$(1, 1, \cdots, 1)^T$.
This basis gives rise to the invariant above, which defines the map
from the weight lattice to ${\mathbb Z}_k$.

Putting this together, the mirror to a pure (supersymmetric) 
$SU(k)/{\mathbb Z}_k$
gauge theory with any choice of discrete theta angle is defined by the
superpotential
\begin{eqnarray}
W & = &
(1/k) \tilde{\sigma}_1 \bigl[  - Z_{12} - Z_{13} - \cdots - Z_{1 (k-1)}
- 2 Z_{1k} - Z_{2k} - \cdots - Z_{(k-1)k}
\nonumber \\
& & \hspace*{0.75in}
+ Z_{21} + Z_{31} + \cdots + Z_{(k-1)1} + 2 Z_{k1} +
Z_{k2} + \cdots + Z_{k(k-1)} \bigr]
\nonumber \\
& & \: + \:
(1/k) \tilde{\sigma}_2 \bigl[ - Z_{21} - Z_{23} - Z_{24} - \cdots -
Z_{2(k-1)} - 2 Z_{2k} - Z_{1k} - Z_{3k} - \cdots - Z_{(k-1)k}
\nonumber \\
& & \hspace*{0.5in}
+ Z_{12} + Z_{32} + Z_{42} + \cdots + Z_{(k-1) 2} + 2 Z_{k2}
+ Z_{k1} + Z_{k3} + Z_{k4} + \cdots + Z_{k(k-1)} \bigr]
\nonumber \\
& & \: + \: \cdots \: + \:
\sum_{a=1}^{k-1} t_a \tilde{\sigma}_a
 \: + \: \sum_{\mu \neq \nu} X_{\mu \nu},
\end{eqnarray}
where
\begin{equation}
t_1, t_2, \cdots, t_{k-1} \: \in 
\{ 0, 2 \pi i/k, 4 \pi i/k, \cdots, (k-1) 2 \pi i/k \},
\end{equation}

It is straightforward to see that the critical locus equations take
the form
\begin{equation}
X_{\mu \nu} \: = \: - X_{\nu \mu},
\end{equation}
and
\begin{eqnarray}
\left[ \frac{ X_{12} }{X_{21}} \frac{X_{13}}{X_{31}} \cdots
\frac{ X_{1(k-1)} }{X_{(k-1)1}} \left( \frac{ X_{1k} }{X_{k1}} \right)^2
\frac{ X_{2k} }{ X_{k2} } \cdots \frac{ X_{(k-1)k} }{ X_{k(k-1)}}
\right]^{1/k}
& = & \exp(-t_1),
\\
\left[ \frac{ X_{21} }{X_{12}} \frac{X_{23}}{X_{32}} \frac{X_{24}}{X_{42}}
\cdots \frac{X_{2(k-1)}}{X_{(k-1)2}} \left( \frac{ X_{2k}}{X_{k2}} \right)^2
\frac{ X_{1k} }{X_{k1}} \frac{X_{3k}}{X_{k3}} \cdots 
\frac{X_{(k-1)k}}{X_{k(k-1)}} 
\right]^{1/k}
& = & \exp(-t_2),
\end{eqnarray}
and so forth.  Each of these equations involves a product of $2(k-1)$ ratios
of $X$'s associated to positive and negative roots. 
 
As in the $SU(3)$ case, it is straightforward to check that branch cuts
change the $t_a$ but not the discrete theta angle.
The most efficient general argument is to observe that
\begin{equation}  \label{eq:suk:genlprod}
\prod_a \exp(-t_a) \: = \: \frac{X_{1k}}{X_{k1}}
\frac{ X_{2k} }{X_{k2}} \cdots \frac{X_{(k-1)k}}{X_{k(k-1)}},
\end{equation}
making it explicit that there is no branch cut ambiguity in
\begin{equation}
\exp\left( \sum_a t_a \right),
\end{equation}
hence the discrete theta angle is well-defined.
(In principle, this is closely related to the fact that
the discrete theta angle is well-defined under root lattice shifts,
realized as log branch cuts in the mirror superpotential.)

Now, let us find the value(s) of the discrete theta angle for which the critical
locus admits solutions.  From equation~(\ref{eq:suk:genlprod}), 
and the fact that on the critical locus each ratio is $-1$, we find 
that the unique discrete theta angle for which a solution exists is
determined by
\begin{equation}
\exp\left( - \sum_a t_a \right) \: = \: 
(-)^{k-1},
\end{equation}
or equivalently,
\begin{equation}
\sum_a m_a \: = \: - \frac{1}{2} k (k-1) \mod k.
\end{equation}
Thus, a pure $SU(k)/{\mathbb Z}_k$ gauge theory will have supersymmetric
vacua for one value of the discrete theta angle, given by
\begin{equation}
- \frac{1}{2} k (k-1) \mod k.
\end{equation}

For one example, for $SO(3) = SU(2)/{\mathbb Z}_2$, this is 
$-1 \mod 2$, and so we confirm that the pure $SO(3)$ gauge
theory has supersymmetric vacua for the one nontrivial choice of
discrete theta angle, matching the results of \cite[section 12.1]{Gu:2018fpm}.

For another example, for $SU(3)/{\mathbb Z}_3$, this is
$-3 \equiv 0 \mod 3$, and so the pure $SU(3)/{\mathbb Z}_3$ gauge theory
has vacua for only the vanishing discrete theta angle,
correctly reproducing our result from section~\ref{sect:su3}.

For later use, for $SU(4)/{\mathbb Z}_4$, this is
$-6 \equiv 2 \mod 4$, and so only for this discrete theta angle
does the pure $SU(4)/{\mathbb Z}_4$ theory have vacua.  
(We will later see the same result for the pure
$SO(6)/{\mathbb Z}_2$ theory, as one should expect since the two
groups are the same.)

Consider that one choice of discrete theta angle for which the
critical locus equations have solutions.
Of the fields $X_{\mu \nu}$ for $\mu < \nu$,
since they are all determined by $k-1$ $\tilde{\sigma}$'s,
there can be only $k-1$ independent $X_{\mu \nu}$'s.  Integrating out
the $\tilde{\sigma}$'s results in a vanishing superpotential,
so we find that the mirror is (plausibly) consistent with
$k-1$ free superfields.

For completeness, let us also verify that the same result can be
obtained from equation~(\ref{eq:gauge:t-crit}).  
Taking the positive roots to be those associated with $Z_{\mu \nu}$ for
$\mu < \nu$, and our previous expressions for those 
in equations~(\ref{eq:suk:z12}) through (\ref{eq:suk:z1k}),
it is straightforward to demonstrate that
\begin{equation}
\sum_{\rm pos'} \alpha^a_{\mu} \: = \: -2k + 2a,
\end{equation}
hence
\begin{equation}
t_a \: = \: - \frac{\pi i}{k} \sum_{\rm pos'} \alpha^a_{\mu}
\: = \:  \frac{2 \pi i}{k} ( k - a ),
\end{equation}
or equivalently
\begin{equation}
m_a \: = \: k-a.
\end{equation}
For the discrete theta angle, we compute
\begin{equation}
\sum_a m_a \: = \: \sum_{a=1}^{k-1} (k-a) \: = \:
+ \frac{1}{2} k (k-1) \: \equiv \: - \frac{1}{2} k (k-1) \mod k,
\end{equation}
which matches the result obtained above for the discrete theta angle.

As a consistency check, we apply decomposition 
\cite{Hellerman:2006zs,Sharpe:2014tca,Sharpe:2019ddn},
which states that two-dimensional theories with one-form symmetries
decompose.  In this case, it implies that a pure $SU(k)$ gauge theory
decomposes into a disjoint union of $k$ $SU(k)/{\mathbb Z}_k$
gauge theories, one with every possible discrete theta angle,
or schematically,
\begin{equation}
SU(k) \: = \: \bigoplus_{n=0}^{k-1} \left( \frac{ SU(k) }{ {\mathbb Z}_k }
\right)_n.
\end{equation}
Since the pure $SU(k)$ theory is believed to flow in the IR to
a set of $k-1$ free twisted chiral superfields
\cite{Aharony:2016jki,Gu:2018fpm},
the disjoint union of pure $SU(k)/{\mathbb Z}_k$ gauge theories
should also flow in the IR to $k-1$ free twisted chiral superfields.
We have argued above that most of the $SU(k)/{\mathbb Z}_k$ theories
have no vacua (and so break supersymmetry in the IR); 
for the single $SU(k)/{\mathbb Z}_k$ theory (with the
right choice of discrete theta angle) for which vacua exist,
we have seen evidence that it flows to $k-1$ (plausibly free) twisted chiral
superfields.  Thus, a disjoint union of $k$ $SU(k)/{\mathbb Z}_k$ 
gauge theories with every possible discrete theta angle appears to have
the same IR limit as the $SU(k)$ gauge theory, as expected from
decomposition.

\subsection{$SU(4)/{\mathbb Z}_2$}
\label{sect:su4:z2}

So far we have considered quotients of $SU(k)$ by all of its
center, but very similar considerations also apply when quotienting
by subsets of the center.  To illustrate the details, and also to provide
a consistency check, we will work through one example, namely
$SU(4)/{\mathbb Z}_2$.

Following our previous analysis, the mirror superpotential
will multiply the $\sigma Z$ terms by a factor of $1/| {\mathbb Z}_2 | = 
1/2$, and the FI parameters $t_a \in \{0, \pi i\}$.
The mirror superpotential then takes the form
\begin{eqnarray}
W & = &
(1/2) \tilde{\sigma}_1 \bigl[  - Z_{12} - Z_{13} 
- 2 Z_{14} - Z_{24}  - Z_{34}
\nonumber \\
& & \hspace*{0.75in}
+ Z_{21} + Z_{31}  + 2 Z_{41} +
Z_{42}  + Z_{43} \bigr]
\nonumber \\
& & \: + \:
(1/2) \tilde{\sigma}_2 \bigl[ - Z_{21} - Z_{23}  
 - 2 Z_{24} - Z_{14} - Z_{34} 
\nonumber \\
& & \hspace*{0.5in}
+ Z_{12} + Z_{32}  + 2 Z_{42}
+ Z_{41} + Z_{43}  \bigr]
\nonumber \\
& & \: + \: \cdots \: + \: 
\sum_{a=1}^{3} t_a \tilde{\sigma}_a
 \: + \: \sum_{\mu \neq \nu} X_{\mu \nu}.
\end{eqnarray}
If we write
\begin{equation}
t_a \: = \: \frac{2 \pi i}{2} m_a,
\end{equation}
then the discrete theta angle is given as
\begin{equation}
\sum_a m_a \mod 2.
\end{equation}

The critical locus equations are
\begin{eqnarray}
\left[ \frac{ X_{12} }{ X_{21} } \frac{ X_{13} }{ X_{31} }
\left( \frac{ X_{14} }{ X_{41} } \right)^2
\frac{ X_{24} }{ X_{42} }
\frac{ X_{34} }{ X_{43} } \right]^{1/2} & = & \exp(-t_1),
\\
\left[ \frac{ X_{21} }{ X_{12} } \frac{ X_{23} }{ X_{32} }
\left( \frac{ X_{24} }{ X_{42} } \right)^2
\frac{ X_{14} }{ X_{41} } \frac{ X_{34} }{ X_{43} } \right]^{1/2}
& = & \exp(-t_2),
\\
\left[ \frac{ X_{31} }{ X_{31} }  \frac{ X_{32} }{ X_{23} }
\left( \frac{ X_{34} }{ X_{43} } \right)^2
\frac{ X_{14} }{ X_{41} } 
\frac{ X_{24} }{ X_{42} } \right]^{1/2} 
& = & \exp(-t_3).
\end{eqnarray}
Plugging into the discrete theta angle equation, we find
\begin{equation}
\exp( - t_1 - t_2 - t_3 ) \: = \:
\left[ \frac{ X_{14} }{ X_{41} }
\frac{ X_{24} }{ X_{42} } 
\frac{ X_{34} }{ X_{43} } \right]^2,
\end{equation}
which is $+1$ on the critical locus (and which does not have any
square-root branch cut ambiguities).  Thus, we see
pure $SU(4)/{\mathbb Z}_2$ gauge theories have supersymmetric vacua
only for the case of vanishing discrete theta angle.

As a consistency check, recall that $SO(6) = SU(4)/{\mathbb Z}_2$,
and it was argued in \cite[section 13.1]{Gu:2018fpm} that pure
$SO(6)$ gauge theories also only have supersymmetric vacua
for vanishing discrete theta angles, which is consistent with our
computation here.

\section{Pure $SO(2k)/{\mathbb Z}_2$ gauge theories}
\label{sect:so2k}

Pure two-dimensional supersymmetric $SO(2k)$ and $SO(2k+1)$ gauge theories
were considered in \cite[sections 13.1, 13.2]{Gu:2018fpm},
which argued that for the nontrivial discrete theta angle,
they admit vacua, but not for theories with no discrete theta angle.
(Since $\pi_1(SO) = {\mathbb Z}_2$, they admit two possible
discrete theta angles.)  
Now, the groups $SO(2k+1)$ have no center, so there are no further quotients
to consider, but the groups $SO(2k)$ have center ${\mathbb Z}_2$,
and so we can consider gauge theories with gauge group $SO(2k)/{\mathbb Z}_2$.
These groups have
(depending upon $k$) fundamental group either
${\mathbb Z}_4$ or ${\mathbb Z}_2 \times {\mathbb Z}_2$, and hence
$SO(2k)/{\mathbb Z}_2$ gauge theories have discrete theta angles counted
by either
${\mathbb Z}_4$ or ${\mathbb Z}_2 \times {\mathbb Z}_2$.

In this section, we will discuss the mirrors to such theories,
including the realization of the
discrete theta angles in the mirror, and the IR behavior of
pure $SO(2k)/{\mathbb Z}_2$ theories.

\subsection{Basics}

Building on \cite[section 13.1]{Gu:2018fpm},
the mirror superpotential to a pure $SO(2k)/{\mathbb Z}_2$ gauge
theory is
\begin{eqnarray}
W & = &
(1/2) \sum_{a=1}^k \sigma_a \left(
 \sum_{\nu > 2a} \left( Z_{2a, \nu} - Z_{2a-1,\nu} \right)
\: + \: \sum_{\mu < 2a-1}  \left( Z_{\mu,2a} - Z_{\mu,2a-1} \right)
\: - \: 2 t_a \right)
\nonumber \\
& & \hspace*{2in}
\: + \: 
\sum_{\mu < \nu} X_{\mu \nu},   \label{eq:so2k:W}
\end{eqnarray}
where $t_a$ describe the discrete theta angle (in an intricate fashion
we shall detail later),
$X_{\mu \nu} = X_{\nu \mu}^{-1}$, and $X_{\mu \nu}$ are defined for
$\mu, \nu \in \{1, \cdots, 2k\}$, excluding
$X_{2a-1,2a}$.
The critical loci are given by\footnote{
This corrects an error in the critical locus equations
given in our earlier work
\cite[section 13.1]{Gu:2018fpm}.
Although the critical locus equations are slightly different,
the conclusions of \cite[section 13.1]{Gu:2018fpm} are unaffected.
}
\begin{equation}
X_{2a,2b} \: = \: - X_{2a-1,2b-1},
\: \: \:
X_{2a-1,2b} \: = \: - X_{2a,2b-1},
\end{equation}
\begin{equation}
\left(
\prod_{\nu > 2a} \frac{ X_{2a-1,\nu} }{ X_{2a, \nu} } \right)^{1/2}
\left(
\prod_{\mu < 2a-1} \frac{ X_{\mu,2a-1} }{ X_{\mu,2a} } \right)^{1/2}
\: = \: \exp(-t_a).
\end{equation}
The $t_a$ define the possible discrete theta angles, in a fashion we shall
describe after walking through the form of the root and weight lattices
next.

To compute possible vacua, we need a more systematic understanding of
the discrete theta angles and their relation to values of $t_a$,
to which we now turn.

\subsection{Weight and root lattices}

Now, let us consider lattice vectors, following the
abstract considerations of section~\ref{sect:rev}.
In the conventions of \cite[sections 13.1, 13.2]{Gu:2018fpm},
the root lattice is generated by vectors of the form
\begin{equation}
(0, \cdots, 0, \pm 1, 0, \cdots, 0, \pm 1, 0, \cdots 0),
\end{equation}
in a basis defined by the weights of the vector representation.
Points on the root lattice include
\begin{equation}
( \pm 2, 0, \cdots, 0), \: \: \:
(0, \pm 2, 0, \cdots, 0),
\end{equation}
and so forth,
but not
\begin{equation}
( \pm 1, 0, \cdots, 0), \: \: \:
(0, \pm 1, 0, \cdots, 0),
\end{equation}
and so forth.

If we write a root lattice basis in terms of the vector representation
basis above, such as
\begin{equation}
\left[ \begin{array}{rr}
1 & 1 \\
-1 & 1 \end{array} \right]
\end{equation}
for $SO(4)$,
we find a minor puzzle -- this matrix has determinant two,
and its Smith decomposition gives a diagonal matrix with entries
$1$, $2$, rather than the expected center of the universal cover (in this case,
${\mathbb Z}_2 \times {\mathbb Z}_2$).  This is because a basis built
from vector representation weights does not span the weight lattice,
in particular it does not include weights of the spinor representations,
which are of the form
\begin{equation}
(1/2) ( \pm 1, \pm 1, \pm 1, \cdots ).
\end{equation}
In the case of $SO(4)$ above, if we write the root lattice basis
$(1,1)$, $(-1,1)$ in terms of a basis for the complete weight lattice,
including spinor representations, such as $(1/2)(1,1)$, $(1/2)(-1,1)$,
then we find that the new matrix is just
\begin{equation}
\left[ \begin{array}{cc}
2 & 0 \\ 0 & 2 \end{array}
\right],
\end{equation}
reflecting the fact that Spin$(4)$ has center ${\mathbb Z}_2 \times
{\mathbb Z}_2$.

In principle, lattice invariants can be read off from the diagonalizing
matrices in the Smith decomposition of the matrix giving a basis of root
vectors.  For one example, for $SO(6)$, such a matrix of a
basis of root vectors
(in a basis of spinor weights) takes the form
\begin{equation}
A \: = \:
\left[ \begin{array}{rrr}
2 & -1 & -1 \\
0 & 1 & -1 \\
-1 & 1 & 2 \end{array} \right].
\end{equation}
A Smith decomposition is defined by
\begin{equation}
Q^{-1} \: = \:
\left[ \begin{array}{rrr}
1 & 0 & 1 \\
1 & 0 & 2 \\
1 & -1 & 2 \end{array} \right],
\: \: \:
P \: = \: \left[ \begin{array}{rrr}
1 & 0 & -1 \\
0 & 1 & -3 \\
0 & 0 & 1
\end{array} \right],
\end{equation}
where
\begin{equation}
\det Q^{-1} \: = \: 1 \: = \: \det P,
\end{equation}
and
\begin{equation}
Q^{-1} A P \: = \: {\rm diag}( 1, 1, 4),
\end{equation}
reflecting the fact that the center of Spin$(4)$ is ${\mathbb Z}_4$.
The last column of $P$ indicates that an invariant function can be
obtained from
\begin{equation}
\vec{\gamma} \cdot \vec{n} \mod 4,
\end{equation}
where in the basis of spinor weights,
\begin{equation}
\vec{\gamma} \: = \: (-1, -3, +1).
\end{equation}
Converting from the basis of spinor weights
\begin{equation}
(1/2)(1, 1, 1), \: \: \:
(1/2) (-1, 1, 1),  \: \: \:
(1/2) (1, -1, 1)
\end{equation}
to the original basis, we have\footnote{
It may be useful to observe that the coordinate change is
\begin{equation}
n_1 \: = \: y_1 - y_2, \: \: \:
n_2 \: = \: y_1 - y_3, \: \: \:
n_3 \: = \: y_2 + y_3.
\end{equation}
}
\begin{equation}
\vec{\gamma} \: = \: (2, -2, -2),
\end{equation}
which
gives the invariant
\begin{equation}
\vec{\gamma} \cdot \vec{n} \: = \:
2 n_1 - 2 n_2 - 2 n_3 \mod 4
\end{equation}
in conventions in which $n_a \in (1/2) {\mathbb Z}$.  If we rescale
to integer-valued vectors $\vec{m}$, this becomes
\begin{equation}
m_1 - m_2 - m_3 \: \equiv \:
\sum_a m_a \mod 4,
\end{equation}
with a basis for the root lattice now of the form
\begin{equation}
(2, 2, 0), \: \: \: (-2, 2, 0), \: \: \:
(0, -2, 2).
\end{equation}
We will see in the next section on different grounds that 
such a sum defines the ${\mathbb Z}_4$-valued discrete theta angle for
mirrors to $SO(4k+2)$ gauge theories.

For one more example, consider $SO(8)$.  A basis of root vectors,
in a basis of spinor weights, is encoded in the rows of the matrix
\begin{equation}
A \: = \: \left[ \begin{array}{rrrr}
2 & -1 & -1 & 0 \\
0 & 1 & -1 & 0 \\
0 & 0 & 1 & -1 \\
-2 & 1 & 1 & 2 \end{array} \right].
\end{equation}
This has a Smith decomposition given by
\begin{equation}
Q^{-1} \: = \: \left[ \begin{array}{rrrr}
1 & 0 & 1 & 1 \\
0 & 0 & 0 & 1 \\
0 & -1 & 0 & 1 \\
-1 & -1 & -2 & 0 \end{array} \right],
\: \: \:
P \: = \: \left[ \begin{array}{rrrr}
1 & 0 & -1 & 0 \\
0 & 1 & -1 & -1 \\
0 & 0 & 1 & -1 \\
1 & 0 & -1 & 1
\end{array} \right],
\end{equation}
where
\begin{equation}
\det Q^{-1} \: = \: 1 \: = \: \det P,
\end{equation}
and
\begin{equation}
Q^{-1} A P \: = \: {\rm diag}(1, 1, 2, 2),
\end{equation}
as expected since the center of $SO(8)$ is ${\mathbb Z}_2 \times
{\mathbb Z}_2$.  We can read off a pair of ${\mathbb Z}_2$ valued
functions from the last two columns of $P$, given by
\begin{equation}
\vec{\gamma}_1 \cdot \vec{n} \mod 2, \: \: \:
\vec{\gamma}_2 \cdot \vec{n} \mod 2,
\end{equation}
where in the basis of spinor weights,
\begin{equation}
\vec{\gamma}_1 \: = \: (-1, -1, +1, -1), \: \: \:
\vec{\gamma}_2 \: = \: (0, -1, -1, +1).
\end{equation}
Converting from the basis of spinor weights
\begin{equation}
(1/2)(1,1,1,1), \: \: \:
(1/2)(-1, 1, 1, 1), \: \: \:
(1/2)(1, -1, 1, 1), \: \: \:
(1/2)(1, 1, -1, 1)
\end{equation}
to the original basis, we have\footnote{
It may be useful to observe that the coordinate change is
\begin{equation}
n_1 \: = \: y_1 - y_2, \: \: \:
n_2 \: = \: y_1 - y_3, \: \: \:
n_3 \: = \: y_1 - y_4, \: \: \:
n_4 \: = \: -y_1 + y_2 + y_3 + y_4.
\end{equation}
}
\begin{equation}
\vec{\gamma}_1 \: = \: (0, -2, 0, 0), \: \: \:
\vec{\gamma}_2 \: = \: (1, 1, -1, -1)
\end{equation}
which gives the invariants
\begin{equation}
\vec{\gamma}_1 \cdot \vec{n} \: = \: -2 n_2 \mod 2, 
\: \: \:
\vec{\gamma}_2 \cdot \vec{n} \: = \: n_1 + n_2 - n_3 - n_4 \mod 2
\end{equation}
in conventions in which $n_a \in (1/2) {\mathbb Z}$.
If we rescale to integer-valued vectors $\vec{m}$, these become
\begin{equation}
m_2 \mod 2, \: \: \:
\sum_a m_a \mod 4,
\end{equation}
where, because of the structure of the weight lattice,
\begin{equation}
\sum_a m_a \mod 4 \: \in \: \{ 0, 2 \},
\end{equation}
and hence defines a ${\mathbb Z}_2$ invariant.

We will see in the next section on different grounds that such sums
define the ${\mathbb Z}_2 \times {\mathbb Z}_2$-valued discrete theta angle
for mirrors to $SO(4k)$ gauge theories.

\subsection{Discrete theta angles}

In this subsection we shall describe how the $t_a$ encode discrete
theta angles of the $SO(2k)/{\mathbb Z}_2$ gauge theory.
The form of the result depends upon the value of $k$:
\begin{itemize}
\item for $k$ odd, possible discrete theta angles are classified by
${\mathbb Z}_4$,
\item for $k$ even, possible discrete theta angles are classified by
${\mathbb Z}_2 \times {\mathbb Z}_2$.
\end{itemize}
In particular, for gauge groups $SO(4\ell + 2)/{\mathbb Z}_2$, 
there is a single
(${\mathbb Z}_4$) discrete
theta angle, whereas for gauge groups $SO(4\ell)/{\mathbb Z}_2$,
there are a pair of (${\mathbb Z}_2$) discrete theta angles.  
We will discuss each of these
cases.

First, write each
\begin{equation}
t_a \: = \: m_a \frac{2 \pi i}{4}.
\end{equation}
Possible values of $m_a$ are encoded in the weight lattice of the Lie
algebra, from our ansatz in section~\ref{sect:rev}.  
As discussed in the previous subsection, allowed weights on the
weight lattice have one of the following two forms:  either all of the $m_a$
are odd, or all of the $m_a$ are even.  In particular, the weights 
corresponding to spinor representations are 
\begin{equation}
(m_1, m_2, m_3, \cdots) \: = \: 
(\pm 1, \pm 1, \pm 1, \cdots)
\end{equation} 
in this normalization convention, or $(1/2)(\pm 1, \pm 1, \cdots)$
in the conventions of \cite{georgi}.  Similarly, root lattice vectors are
of the form
\begin{equation}
(m_1, m_2, m_3, \cdots) \: = \: (\cdots, 0, \pm 2, 0, \cdots, 0, \pm 2, 0,
\cdots),
\end{equation}
and so shifts by root lattice elements shift pairs of $m_a$ by $\pm 2$.

Given the structure of possible weight lattice elements $(m_a)$ above,
we can now define two functions that are invariant under root lattice shifts:
\begin{equation}
\sum_a m_a \mod 4,  \label{eq:so2k:inv1}
\end{equation}
\begin{equation}
m_1 \mod 2.  \label{eq:so2k:inv2}
\end{equation}
(The reader should note that there is nothing special about $m_1$,
the invariant can be defined for any one $m_a$, and gives the same
value for any choice.)
The second invariant tells whether the weight lattice element is one for
which all of the $m_a$ are odd or even.

Now, in the case that $k$ is even, more can be said.
In this case, for all weight lattice elements,
\begin{equation}
\sum_a m_a \: \equiv \: 0 \mod 2.
\end{equation}
Put another way, the sum
\begin{equation}
\sum_a m_a \: \in \{ 0, 2 \} \mod 4.
\end{equation}
As a result, in this case, the two invariants~(\ref{eq:so2k:inv1}),
(\ref{eq:so2k:inv2}) define two separate ${\mathbb Z}_2$-valued invariants,
and so we can naturally identify discrete theta angles with elements
of ${\mathbb Z}_2 \times {\mathbb Z}_2$.

In the case that $k$ is odd,
the first invariant~(\ref{eq:so2k:inv1}) can take any value (mod 4),
and so defines a map to ${\mathbb Z}_4$.  The second invariant,
(\ref{eq:so2k:inv2}), merely tells us whether the first invariant will
be even or odd, and so does not provide independent information.
Thus, we see that in this case, the possible discrete theta angles
are characterized by ${\mathbb Z}_4$.

Finally,
let us check this prescription for the discrete theta angle is consistent
with the Weyl group.  The Weyl group $W$ of any $SO(2k)$ is
\cite[section 18.1]{fh}
the extension
\begin{equation}
1 \: \longrightarrow \: K \: \longrightarrow \: W \: \longrightarrow \:
S_k \: \longrightarrow \: 1,
\end{equation}
where $K \subset ({\mathbb Z}_2)^k$
is the subgroup with an even number of nontrivial generators.  Each
${\mathbb Z}_2$ factor multiplies the corresponding $\sigma_a$ by
$-1$, and hence also multiplies $t_a$ by $-1$.  (There is also
an action on the $X_{\mu \nu}$, see \cite[section 9]{Gu:2018fpm}
for details.)
Since there are an even
number of sign flips, and each sign flip changes $m_a \mapsto m_a + 2$,
we see that $\sum m_a \mod 4$ is well-defined under $K$, and since it
is permutation-invariant, it is also invariant under all of the 
Weyl group $W$.  Similarly, since all the $m_a$ are either even or odd,
$m_1 \mod 2$ is also permutation-invariant, and as it is a mod 2 invariant,
it is also invariant under $K$ and so invariant under all of the
Weyl group.

\subsection{Vacua}

Now that we understand the discrete theta angles in the mirror,
we can turn to computing the vacua.

Recall that
the critical loci of the mirror superpotential~(\ref{eq:so2k:W})
to a pure $SO(2k)/{\mathbb Z}_2$ gauge theory
are given by
\begin{equation}
X_{2a,2b} \: = \: - X_{2a-1,2b-1},
\: \: \:
X_{2a-1,2b} \: = \: - X_{2a,2b-1},
\end{equation}
\begin{equation}
\left(
\prod_{\nu > 2a} \frac{ X_{2a-1,\nu} }{ X_{2a, \nu} } \right)^{1/2}
\left(
\prod_{\mu < 2a-1} \frac{ X_{\mu,2a-1} }{ X_{\mu,2a} } \right)^{1/2}
\: = \: \exp(-t_a).  \label{eq:so2k:crit2}
\end{equation}

Because of the square root branch cuts, the $t_a$ are not themselves
uniquely defined, so we shall turn to the discrete theta angles,
for which we will be able to get well-defined expressions.
There are two cases, $k$ even or odd, with related invariants,
which we consider in turn.

First, we consider the 
discrete theta angle defined by
\begin{equation}
\sum_a m_a \mod 4,
\end{equation}
so we consider the corresponding product of the $\exp(-t_a)$.
(For $k$ odd, this takes values in ${\mathbb Z}_4$, but for $k$ even,
the sum lies in $\{0, 2\}$, and so for $k$ even this is only 
${\mathbb Z}_2$-valued.)
It is straightforward to show that on the critical locus,
from equation~(\ref{eq:so2k:crit2}),
\begin{equation}
\prod_a \exp(-t_a) \: = \: 
\prod_{a < b} \frac{ X_{2a-1,2b-1} }{ X_{2a,2b} }.
\end{equation}
(Note that the square roots have disappeared -- those factors which
did not cancel out, came in pairs, giving a well-defined expression.)
Furthermore, on the critical locus,
each of the ratios is $-1$, and so we find
\begin{equation}
\prod_a \exp(-t_a) \: = \: (-1)^N,
\end{equation}
where
\begin{equation}
N \: = \: \left( \begin{array}{c} k \\ 2 \end{array} \right)
\: = \: \frac{k (k-1)}{2}.
\end{equation}
Put another way, the discrete theta angle for which the theory has
supersymmetric vacua is 
\begin{equation}
\sum_a m_a \: \equiv \: 2 N \mod 4 \: \equiv \: k (k-1) \mod 4.
\end{equation}
For example, a pure $SO(6)/{\mathbb Z}_2$ gauge theory
can only have vacua for the case that 
\begin{equation}
\sum_a m_a \: \equiv \: 2 \left( \begin{array}{c} 3 \\ 2
\end{array} \right) \mod 4 \: \equiv \: 2 \mod 4.
\end{equation}
(As a consistency check, recall that $SO(6)/{\mathbb Z}_2 =
SU(4)/{\mathbb Z}_4$, and for the latter, we also found that
vacua exist only for the same discrete theta angle 
as above.)

The second invariant is 
\begin{equation}
m_1 \mod 2.
\end{equation}
For $k$ odd, this merely defines the cokernel of the map
${\mathbb Z}_2 \hookrightarrow {\mathbb Z}_4$ 
but for $k$ even, this defines a distinct independent ${\mathbb Z}_2$ invariant.
Since the $t_a \in (\pi i/2) {\mathbb Z}$, to get this invariant we consider
\begin{equation}
\exp(-2 t_1) \: = \: \prod_{\nu > 2} \frac{ X_{1 \nu} }{ X_{2 \nu} }
\: = \: \prod_{a > 1} \frac{ X_{1, 2a} }{ X_{2, 2a-1} } 
\frac{ X_{1, 2a-1} }{ X_{2, 2a} }.
\end{equation}
On the critical locus,
each ratio is $-1$, and there are
$2k-2$ such factors, hence
\begin{equation}
\exp(-2 t_1) \: = \: +1,
\end{equation}
and so
\begin{equation}
m_1 \: \equiv \: 0 \mod 2.
\end{equation}
and for the case $k$ is even, this is an independent
${\mathbb Z}_2$ invariant.

So far, we have computed discrete theta angles for which
pure $SO(2k)/{\mathbb Z}_2$ gauge theories 
admit vacua.  (There is a unique choice of discrete theta angle
for which the theory has supersymmetric vacua, meaning there are no
supersymmetric vacua for different discrete theta angles.)
For the case when there are vacua, the same arguments that have appeared
elsewhere suggest that the IR limit is a theory of free twisted chiral
superfields, as many as the rank.

This is consistent with decomposition
\cite{Hellerman:2006zs,Sharpe:2014tca,Sharpe:2019ddn}:  a pure $SO(2k)$
gauge theory admits a ${\mathbb Z}_2$ one-form symmetry, and decomposes into
a pair of $SO(2k)/{\mathbb Z}_2$ gauge theories, with two of the possible
four discrete theta angles, for example
\begin{equation}
SO(2k)_0 \: = \: \left( SO(2k)/{\mathbb Z}_2 \right)_0
+
\left( SO(2k)/{\mathbb Z}_2 \right)_2
\end{equation}
(where subscripts on the right indicate $\sum_a m_a$, mod $4$).
We argued in \cite[section 13.1]{Gu:2018fpm} that the pure $SO(2k)$ gauge
theory only has supersymmetric vacua for the case of vanishing discrete
theta angle, and we have seen above that pure $SO(2k)/{\mathbb Z}_2$
gauge theories only have supersymmetric vacua for one of two discrete
theta angles, for which $\sum_a m_a \in \{0, 2\}$.  Precisely those
two $SO(2k)/{\mathbb Z}_2$ gauge theories appear in the decomposition
of $SO(2k)_0$, and as the structure of the vacua is the
same in both cases, we see that our results are consistent with
decomposition.

\subsection{Pure $SO(2k)$ gauge theories}

Now that we have given a complete picture of discrete theta angles
in $SO(2k)/{\mathbb Z}_2$ gauge theories, let us re-examine the picture
of discrete theta angles in $SO(2k)$ gauge theories given in
\cite{Gu:2018fpm}.  There, instead of multiple $t_a$, only a single
$t$ was specified, taking values in $\{ 0, \pi i\}$.
Although this description is rather
different from the more sophisticated
description of discrete theta angles used elsewhere in this paper,
we shall see that it is appropriate for that case.

In principle, we can derive discrete theta angles for $SO(2k)$
gauge theories using the same methods we have applied elsewhere
in this paper.  In principle, we have a set of
$t_a$, as many as the rank, and since the discrete theta angles
here take values in ${\mathbb Z}_2$, each $t_a \in \{0, \pi i \}$.
As before, write
\begin{equation}
t_a \: = \: \frac{2 \pi i}{2} m_a,
\end{equation}
for $m_a \in {\mathbb Z}$.

As before, the weight lattice consists of sets of integers $(m_a)$ that
are either all even or all odd.  Since we are only interested in the
integers mod two, the all-even case is equivalent to taking all of the
$m_a$ to be zero, and the all-odd case is equivalent to taking all of
the $m_a$ to be one.  In either case, note that all of the $m_a$
are identical, hence all of the $t_a$ are identical, agreeing with
the description in \cite{Gu:2018fpm}.  Furthermore, the discrete theta
angle can be encoded as
\begin{equation}
m_1 \mod 2,
\end{equation}
which is equivalent to the description in \cite{Gu:2018fpm}.

Next, let us check that our results are consistent with decomposition
of pure $SO(2k)$ gauge theories
\cite{Hellerman:2006zs,Sharpe:2014tca,Sharpe:2019ddn}.
First, consider the case that $k$ is even, $k=2k'$.
The original $SO(2k) = SO(4k')$ gauge theory has a ${\mathbb Z}_2$
discrete theta angle, and denoting ${\mathbb Z}_2$ discrete theta angles by
a subscript $\pm$, decomposition predicts\footnote{
In fact, we can see this structure from the mirror.
In the mirror to the pure $SO(4k')$ theory, the $+$ discrete theta angle
corresponds to the $m_a$ all even, and the $-$ to the $m_a$ all odd.
In the mirror to the pure $SO(4k')/{\mathbb Z}_2$ theory,
the $++$ case corresponds to all-even $m_a$ with a sum that vanishes mod 4,
the $--$ case corresponds to all-odd $m_a$ with a sum that is 2 mod 4,
the $+-$ case corresponds to all-even $m_a$ with a sum that is 2 mod 4,
and the $-+$ case corresponds to all-odd $m_a$ with a sum that vanishes mod 4.
The terms are simply being grouped by the common ${\mathbb Z}_2$
invariant.
}
\begin{eqnarray}
SO(4k')_+ & = & ( SO(4k')/{\mathbb Z}_2 )_{++} \: \oplus \:
( SO(4k')/{\mathbb Z}_2 )_{+-},
\\
SO(4k')_- & = & ( SO(4k')/{\mathbb Z}_2 )_{--} \: \oplus \:
( SO(4k')/{\mathbb Z}_2 )_{-+}.
\end{eqnarray}
In \cite[section 13.1]{Gu:2018fpm}, it was argued that
the pure $SO(4k')_+$ theory (with vanishing discrete theta angle)
flows in the IR to $2k'$ twisted chiral multiplets,
whereas the $SO(4k')_-$ theory has no supersymmetric vacua.
In this section we have argued that only for either the $++$ discrete theta
angle or the $+-$ discrete theta angle (depending upon $k'$)
does the pure $SO(4k')/{\mathbb Z}_2$ theory admit supersymmetric
vacua, in which case it flows in the IR to $2k'$ twisted chiral multiplets.
We see from the statement of decomposition above that the decomposition of the
$SO(4k')_+$ theory (with the free IR endpoint) includes an
$(SO(4k')/{\mathbb Z}_2)_{++}$ summand, so both sides of the equalities
above have the same IR endpoints, as expected.

Next, consider the case that $k$ is odd, $k=2k'+1$.
The original $SO(4k'+2)$ gauge theory has a ${\mathbb Z}_2$ discrete
theta angle, and only for the case that the discrete theta angle
is trivial 
does the theory have supersymmetric vacua (for which case, it flows in the
IR to a set of twisted chiral multiplets).
Decomposition predicts
\begin{eqnarray}
SO(4k'+2)_+ & = & ( SO(4k'+2)/{\mathbb Z}_2 )_{0} \: \oplus \:
( SO(4k'+2)/{\mathbb Z}_2 )_{2},
\\
SO(4k'+2)_- & = & ( SO(4k'+2)/{\mathbb Z}_2 )_1 \: \oplus \:
( SO(4k'+2)/{\mathbb Z}_2 )_3.
\end{eqnarray}
and we argued earlier that only for either the $0$ discrete theta angle 
or the $2$ discrete theta angle (depending upon $k'$) does
the pure $SO(4k'+2)/{\mathbb Z}_2$ gauge theory admit supersymmetric
vacua.  Thus, both sides of each equality above flow in the IR to the
same endpoint, as expected.

\section{Pure $Sp(2k)/{\mathbb Z}_2$ gauge theories}
\label{sect:sp2k}

The IR behavior of pure supersymmetric two-dimensional
$Sp(2k)$ gauge theories was discussed with nonabelian mirrors in
\cite[section 13.3]{Gu:2018fpm}, using nonabelian mirrors.
In this section we will discuss the IR behavior of pure 
$Sp(2k)/{\mathbb Z}_2$ gauge theories.

The center of $Sp(2k)$ is ${\mathbb Z}_2$, hence we can consider
pure gauge theories with gauge group $Sp(2k)/{\mathbb Z}_2$, which have
two possible discrete theta angles.

Drawing upon the mirror to pure $Sp(2k)$ gauge theories discussed
in \cite[section 13.3]{Gu:2018fpm}, the mirror superpotential is given by
\begin{eqnarray}
W & = &
(1/2) \sum_{a=1}^k \sigma_a \Biggl(
 \sum_{\mu \leq \nu}  \left( \delta_{\mu, 2a} 
 - \delta_{\mu, 2a-1} +
\delta_{\nu, 2a} - \delta_{\nu, 2a-1}\right) Z_{\mu \nu} 
\: - \: 2t_a \Biggr)
\nonumber \\
& & 
\: + \: \sum_{\mu} X_{\mu \mu} \: + \: 
\sum_{a < b} \left( X_{2a,2b} + X_{2a-1,2b-1} +
X_{2a-1,2b} + X_{2a,2b-1} \right),
\end{eqnarray}
where $t_a \in \{0, \pi i\}$ encodes the discrete theta angle,
as we shall elaborate momentarily,
and the $X_{\mu \nu}$ are defined for $\mu \leq \nu$, excluding
$X_{2a-1,2a}$.
The critical locus is given by
\begin{equation}
X_{2a,2b} \: = \: - X_{2a-1,2b-1},
\: \: \:
X_{2a-1, 2b} \: = \: - X_{2a,2b-1},
\end{equation}
\begin{equation}  \label{eq:sp2k2:crit}
\left( \frac{ X_{2a-1,2a-1} }{ X_{2a,2a} } \right)
\left( \prod_{b > a} \frac{ X_{2a-1,2b-1} }{ X_{2a,2b} }
\frac{ X_{2a-1,2b} }{ X_{2a,2b-1} } \right)^{1/2}
\left( \prod_{b<a} \frac{ X_{2b-1,2a-1} }{ X_{2b,2a} }
\frac{ X_{2b,2a-1} }{ X_{2b-1,2a} } \right)^{1/2} \: = \: \exp(-t_a).
\end{equation}
To classify possible solutions, we need first to understand the
discrete theta angles.

Write $t_a = m_a \pi i$, for $m_a \in \{0, 1\}$,
then the discrete theta angle is given by
\begin{equation}  \label{eq:sp2k2:dtheta}
\sum_a m_a \mod 2.
\end{equation}
This expression is essentially fixed by the Weyl group.
(We will also give a systematic derivation from weight and root lattices
momentarily.)
Recall that
the Weyl group $W$ of $Sp(2k)$ is \cite[section 16.1]{fh} an extension
\begin{equation}
1 \: \longrightarrow \: ({\mathbb Z}_2)^k \: \longrightarrow \: W
\: \longrightarrow \: S_k \: \longrightarrow \: 1.
\end{equation}
The $({\mathbb Z}_2)^k$ acts on the $\sigma_a$ (and hence $t_a$) by sign
flips, and $S_k$ interchanges the $\sigma_a$ (and $t_a$).
(There is also an action on $X_{\mu \nu}$, as discussed
in \cite[section 11]{Gu:2018fpm}.)
Note that since $t = \pi i$ is equivalent to $t = - \pi i$,
invariance under the $({\mathbb Z}_2)^k$ fixes allowed
$t_a$ to be either zero or $\pi i$, and invariance under
permutations requires that the discrete theta angle be determined
by a sum of $m_a$ (up to irrelevant signs), mod $2$.  Thus,
the discrete theta angle given in expression~(\ref{eq:sp2k2:dtheta})
is invariant under the Weyl group, and in fact is fixed by
Weyl invariance.

Now, let us determine for which discrete theta angles there are
supersymmetric vacua.  From equation~(\ref{eq:sp2k2:crit}),
it is straightforward to show that
\begin{equation}
\prod_a \exp(-t_a) \: = \:
\left( \prod_a \frac{ X_{2a-1,2a-1} }{ X_{2a,2a} } \right)
\left( \prod_{a < b} \frac{ X_{2a-1,2b-1} }{ X_{2a,2b} } \right),
\end{equation}
an expression which does not have any square root branch cuts and hence
is unambiguous.  Each of the ratios appearing above is $-1$ on the critical
locus, 
and since there are
\begin{equation}
k \: + \: \left( \begin{array}{c} k \\ 2 \end{array} \right) \: = \:
\frac{1}{2} k (k+1)
\end{equation}
factors, we see that a necessary condition for the existence of
supersymmetric vacua is that
\begin{equation}
\sum_a m_a \mod 2 \: \equiv \: \frac{1}{2} k (k+1) \mod 2.
\end{equation}
Only for that discrete theta angle does the pure $Sp(2k)/{\mathbb Z}_2$
theory admit supersymmetric vacua.

As a consistency check, for the special case
\begin{equation}
SO(3) \: = \: SU(2) / {\mathbb Z}_2 \: = \:
Sp(2) / {\mathbb Z}_2,
\end{equation}
we see the discrete theta angle is nontrivial (since $(1/2)k(k+1) \equiv
1 \mod 2$), which agrees with the results of
section~\ref{sect:su2}.

As another consistency check, for the special case
\begin{equation}
SO(5) \: = \: Sp(4)/{\mathbb Z}_2,
\end{equation}
to get supersymmetric vacua, the discrete theta angle must be
$(1/2)k(k+1) \equiv 1 \mod 2$,
which agrees with results for $SO(5)$ in 
\cite[section 13.2]{Gu:2018fpm}.

Next, let us turn to weight and root lattices, to confirm our
description of the discrete theta angle, and also help set up
the more complex $E_6$ and $E_7$ cases we will discuss later.  
Following the conventions
and notation of \cite{Gu:2018fpm}, for $Sp(2k)$, the weight lattice
can be identified with ${\mathbb Z}^k$, with the standard basis.
Root vectors are of the form
\begin{equation}
(0, \cdots, \pm 2, \cdots, 0), \: \: \:
(0, \cdots, 0, \pm 1, 0, \cdots, 0, \pm 1, 0, \cdots).
\end{equation}
We can take a basis for the root lattice to be the vectors forming the
rows of the matrix
\begin{equation}
A \: = \: \left[ \begin{array}{rrrrrrr}
1 & 1 & 0 & 0 & \cdots & 0 & 0 \\
1 & -1 & 0 & 0 & \cdots & 0 & 0 \\
0 & 1 & -1 & 0 & \cdots & 0 & 0 \\
0 & 0 & 1 & -1 & \cdots & 0 & 0\\
\vdots & & & & \cdots &  & 0\\
0 & 0 & 0 & 0 & \cdots & 1 & -1
\end{array} \right].
\end{equation}
A Smith decomposition of $A$ is given by
\begin{equation}
Q^{-1} \: = \: \left[ \begin{array}{rrrrrr}
1 & 0 & -1 & -1 & \cdots & -1 \\
1 & -1 & -1 & -1 & \cdots & -1 \\
1 & -1 & -2 & -1 & \cdots & -1 \\
1 & -1 & -2 & -2 & \cdots & -1 \\
\vdots & & & & \cdots & -1 \\
1 & -1 & -2 & -2 & \cdots & -2 
\end{array} \right],
\: \: \:
P \: = \: \left[ \begin{array}{rrrrrr}
1 & 0 & 0 & 0 & \cdots & -1 \\
0 & 1 & 0 & 0 & \cdots & -1 \\
0 & 0 & 1 & 0 & \cdots & -1 \\
0 & 0 & 0 & 1 & \cdots & -1 \\
\vdots & & & & \cdots & -1 \\
0 & 0 & 0 & 0 & \cdots & +1
\end{array} \right].
\end{equation}
It is straightforward to check that
\begin{equation}
\det Q^{-1} \: = \: \pm 1, \: \: \:
\det P \: = \: 1,
\end{equation}
and
\begin{equation}
Q^{-1} A P \: = \: {\rm diag} (1, 1, \cdots, 1, 2),
\end{equation}
as expected since the center of $Sp(2k)$ is ${\mathbb Z}_2$.

We can construct an invariant of the weight lattice, a map
to ${\mathbb Z}_2$, as
\begin{equation}
\vec{\gamma} \cdot \vec{m} \mod 2,
\end{equation}
where $\vec{\gamma}$ is determined from the matrix $P$ to be
\begin{equation}
\gamma \: = \: (-1,-1,-1, \cdots, -1, +1).
\end{equation}
In particular,
\begin{equation}
\vec{\gamma} \cdot \vec{m} \: \equiv \: \sum_a m_a \mod 2,
\end{equation}
agreeing with the expression given earlier for the discrete theta
angle.

So far we have derived for which discrete theta angles a pure 
$Sp(2k)/{\mathbb Z}_2$ gauge theory has supersymmetric vacua.
Following the same analysis as \cite[section 13.3]{Gu:2018fpm},
one can integrate out the $X$ fields to find a theory of 
(Weyl-orbifold-invariant combinations of) $k$ $\sigma$ fields,
with vanishing superpotential.  As elsewhere in this paper, that does
not suffice to uniquely specify that the IR theory is a free theory,
but it is certainly consistent with that possibility.

Let us conclude this section by briefly commenting on decomposition
\cite{Hellerman:2006zs,Sharpe:2014tca,Sharpe:2019ddn},
which in this case says that a pure two-dimensional $Sp(2k)$ gauge
theory can be written as the disjoint union of a pair of
$Sp(2k)/{\mathbb Z}_2$ gauge theories, one for each choice of discrete
theta angle.  We have learned that precisely one of those
pure $Sp(2k)/{\mathbb Z}_2$ gauge theories flows in the IR to the
same endpoint as the pure $Sp(2k)$ gauge theory, as discussed
in \cite[section 13.3]{Gu:2018fpm}.
The other $Sp(2k)/{\mathbb Z}_2$ gauge theory has no supersymmetric
vacua, breaking supersymmetry in the IR.
Thus, we see that these computations are consistent
with the prediction of decomposition.

\section{Pure $E_6/{\mathbb Z}_3$ gauge theories}
\label{sect:e6}

Next, we turn to pure gauge theories with exceptional gauge groups.
Pure gauge theories with gauge groups $G_2$, $F_4$, $E_{6,7,8}$ were
considered in \cite{Chen:2018wep}.  Two of these groups -- 
$E_6$ and $E_7$ -- have a nontrivial center, but only those two
\cite[appendix A]{Distler:2007av}.  Specifically, $E_6$ has center
${\mathbb Z}_3$ and $E_7$ has center ${\mathbb Z}_2$.

In this section we will consider the case $E_6/{\mathbb Z}_3$.
As the pertinent equations are extremely lengthy, we shall be brief,
referring to \cite[section 5]{Chen:2018wep} for details.

Following the same method as before, the mirror superpotential to a pure
$E_6/{\mathbb Z}_3$ gauge theory is of
the form
\begin{equation}
W \: = \: (1/3) \sum_{a=1}^6 \sigma_a \sum_{\tilde{\mu}} \alpha^a_{\tilde{\mu}}
Z_{\tilde{\mu}} \: - \: \sum_a t_a \sigma_a \: + \:
\sum_{\tilde{\mu}} X_{\tilde{\mu}},
\end{equation}
where the $t_a$ determine discrete theta angles, in a fashion to be discussed
briefly.

Briefly, using the results of \cite{Chen:2018wep},
the critical locus equations for the mirror superpotential
are of the form
\begin{equation} \label{eq:e6:crit}
\left[ R_a \right]^{1/3} \: = \: \exp(-t_a),
\end{equation}
where $a \in \{1, \cdots, 6\}$, and each $R_a$ is a product of
$22$ ratios (including multiplicities), in which on the critical locus
each ratio is $-1$.  Because of the cube root, the $t_a$ themselves
are ambiguous.  To determine the necessary conditions for supersymmetric
vacua to exist, we first need to find an expression for discrete theta
angles, and then using that expression, we will find an expression
for the discrete theta angle arising on the critical locus.

Now, to finish classifying solutions, we need to understand how the
possible $t_a$ correspond to the three possible discrete theta angles.
Unlike our previous examples, this cannot be done merely by inspection,
as the lattice is more complicated.  To understand the discrete theta
angles, we need to work more systematically.

Thus, we now turn to weight and root lattices, to understand how the
$t_a$ correspond to
possible discrete theta angles.
Following the conventions and notation of \cite{Chen:2018wep},
a basis for the weight lattice can be given as
\begin{eqnarray}
Y_1 & \sim & (1, 0, 0, 0, 0, 0),
\\
Y_2 & \sim & (-1, 1, 0, 0, 0, 0),
\\
Y_3 & \sim & (0, -1, 1, 0, 0, 0),
\\
Y_{25} & \sim & (0, 0, -1, 1, 0, 0),
\\
Y_{26} & \sim & (0, 0, 0, -1, 1, 0),
\\
Y_7 & \sim & (0, 0, 0, 0, -1, 1).
\end{eqnarray}
We can then write a basis of root lattice vectors as
\begin{eqnarray}
X_1 & \sim & (0, 0, 0, 0, 0, 1) \: = \:
Y_1 + Y_2 + Y_3 + Y_{25} + Y_{26} + Y_7,
\\
X_2 & \sim & (0, 0, 1, 0, 0, -1) \: = \:
- Y_{25} - Y_{26} - Y_7,
\\
X_3 & \sim & (0, 1, -1, 1, 0, 0) \: = \:
Y_1 + Y_2 + Y_{25},
\\
X_4 & \sim & (0, 1, 0, -1, 1, 0) \: = \:
Y_1 + Y_2 + Y_{26},
\\
X_5 & \sim & (1, -1, 0, 1, 0, 0) \: = \:
Y_1 + Y_3 + Y_{25},
\\
X_6 & \sim & (-1, 0, 0, 1, 0, 0) \: = \:
Y_2 + Y_3 + Y_{25},
\end{eqnarray}
which can be encoded in the matrix
\begin{equation}
A \: = \: \left[ \begin{array}{rrrrrr}
1 & 1 & 1 & 1 & 1 & 1 \\
0 & 0 & 0 & -1 & -1 & -1 \\
1 & 1 & 0 & 1 & 0 & 0 \\
1 & 1 & 0 & 0 & 1 & 0 \\
1 & 0 & 1 & 1 & 0 & 0 \\
0 & 1 & 1 & 1 & 0 & 0
\end{array} \right].
\end{equation}
A Smith decomposition of $A$ is given by
\begin{equation}
Q^{-1} \: = \:
\left[ \begin{array}{rrrrrr}
3 & 1 & 1 & -2 & -1 & -2 \\
3 & 1 & 1 & -2 & -2 & -1 \\
3 & 1 & 0 & -2 & -1 & -1 \\
2 & 0 & 1 & -2 & -1 & -1 \\
2 & 0 & 0 & -1 & -1 & -1 \\
4 & 1 & 1 & -3 & -2 & -2
\end{array} \right],
\: \: \:
P \: = \: \left[ \begin{array}{rrrrrr}
1 & 0 & 0 & 0 & 0 & -2 \\
0 & 1 & 0 & 0 & 0 & -2 \\
0 & 0 & 1 & 0 & 0 & -2 \\
0 & 0 & 0 & 1 & 0 & -2 \\
0 & 0 & 0 & 0 & 1 & -2 \\
0 & 0 & 0 & 0 & 0 & 1  \end{array} \right].
\end{equation}
It is straightforward to check that
\begin{equation}
\det Q^{-1} \: = \: 1 \: = \: \det P,
\end{equation}
and
\begin{equation}
Q^{-1} A P \: = \: {\rm diag}(1, 1, 1, 1, 1, 3),
\end{equation}
as expected since the center of $E_6$ is ${\mathbb Z}_3$.
From the description above, we see that the ${\mathbb Z}_3$ invariant is
defined by
\begin{equation}
\vec{\gamma} \cdot \vec{m} \mod 3,
\end{equation}
where $\vec{m} = (m_1, \cdots, m_6)$, and in the basis of $Y$'s,
\begin{equation}
\vec{\gamma} \: = \: (-2, -2, -2, -2, -2, +1).
\end{equation}
Now,
\begin{eqnarray}
y_1 & = & n_1, \\
y_2 & = & - n_1 + n_2, \\
y_3 & = & - n_2 + n_3, \\
y_{25} & = & - n_3 + n_4, \\
y_{26} & = & - n_4 + n_5, \\
y_7 & = & - n_5 + n_6.
\end{eqnarray}
hence
\begin{eqnarray}
n_1 & = & y_1, \\
n_2 & = & y_1 + y_2, \\
n_3 & = & y_1 + y_2 + y_3, \\
n_4 & = & y_1 + y_2 + y_3 + y_{25},\\
n_5 & = & y_1 + y_2 + y_3 + y_{25} + y_{26} , \\
n_6 & = & y_1 + y_2 + y_3 + y_{25} + y_{26} + y_7,
\end{eqnarray}
so in the original basis, we have
\begin{equation}
\vec{\gamma} \: = \: (-2, -4, -6, -8, -10, -9),
\end{equation}
so the predicted invariant is
\begin{equation}
\vec{\gamma} \cdot \vec{m} \: \equiv \: m_1 + 2 m_2 + m_4 + 2 m_5
\mod 3.
\end{equation}
It is straightforward that all the root vectors $\vec{m}$ have
$\vec{\gamma} \cdot \vec{m} \equiv 0 \mod 3$,
in the basis conventions of \cite{Chen:2018wep}.

We take the $t_a$ to be of the form
\begin{equation}
t_a \: = \: \frac{2 \pi i}{3} m_a,
\end{equation}
where $m_a \in {\mathbb Z}$.  
From the weight/root lattice analysis above, we see that the
element of ${\mathbb Z}_3$ defined by a set of $m_a$ is
\begin{equation}
 m_1 + 2 m_2 + m_4 + 2 m_5
\mod 3.
\end{equation}

Now that we have determined the discrete theta angles,
we are ready to compute for which discrete theta angles there are
supersymmetric vacua.  Using the critical locus equations
from \cite[section 5]{Chen:2018wep}, with exponents of $1/3$
as in equation~(\ref{eq:e6:crit}), it is straightforward to compute that
\begin{equation}
\exp(- t_1 - 2 t_2 - t_4 - 2 t_5) \: = \:
\frac{ R_3 R_4 }{R_{10}} \frac{R_{11}}{R_{12}}
\frac{ R_{13} R_{15} R_{17} }{ R_{23} R_{27} R_{28} }
\frac{R_{29} R_{30} R_{31} }{R_{32} } R_{35},
\end{equation}
where each $R_n$ indicates a ratio of the form $X_n/X_{n+36}$,
and all such ratios are $-1$ on the critical locus, from
\cite[section 8]{Chen:2018wep}.  (We should emphasize that the fact
that integer powers of the $R_n$ appear, so that there is no cube root
branch cut ambiguity, is itself a highly nontrivial check,
as the critical locus equations themselves involve factors of the form
$R^{1/3}$.  All factors of $R$ either cancel out or
appear in multiples of three.)  Since there are sixteen such factors,
an even number of factors of $-1$, we see that
\begin{equation}
\exp(- t_1 - 2 t_2 - t_4 - 2 t_5) \: = \: 1
\end{equation}
on the critical locus, so that supersymmetric vacua only exist in the
case
\begin{equation}
m_1 + 2 m_2 + m_4 + 2 m_5 \: \equiv \: 0 \mod 3,
\end{equation}
i.e. for vanishing discrete theta angle.

For that discrete theta angle, as in the analysis in
\cite[section 5.4]{Chen:2018wep} for a pure $E_6$ theory,
the critical locus is described by six twisted chiral multiplets with
vanishing superpotential..

Putting this together, and comparing to our earlier results,
a pure $E_6/{\mathbb Z}_3$ gauge theory has vacua (corresponding to
six twisted chiral multiplets) 
in the case that all $t_a$ vanish, which is the case
of vanishing discrete theta angle.  As before, this is consistent
with decomposition:  a pure $E_6$ gauge theory decomposes into a sum
of pure $E_6/{\mathbb Z}_3$ theories, and precisely one of those
$E_6/{\mathbb Z}_3$ theories has a nontrivial IR limit, the same
as the pure $E_6$ theory, while the others break supersymmetry in the IR.

\section{Pure $E_7/{\mathbb Z}_2$ gauge theories}
\label{sect:e7}

In this section we will consider the case $E_7/{\mathbb Z}_2$.
As the pertinent equations are extremely lengthy, we shall be brief,
referring to \cite[section 6]{Chen:2018wep} for details.

Briefly, the critical locus equations for the mirror superpotential
are of the form
\begin{equation} \label{eq:e7:crit} 
\left[ R_a \right]^{1/2} \: = \: \exp(-t_a),
\end{equation}
where $a \in \{1, \cdots, 7\}$, and each $R_a$ is a product of
$68$ ratios (including multiplicities), in which on the critical locus
each ratio is $-1$.  Because of the square root branch cuts,
the values of $\exp(-t_a)$ are not uniquely determined on the critical
locus; however, we will find a unique value of the discrete theta angle
for which solutions exist.

To complete the analysis, we need to describe how the $t_a$ encode
discrete theta angles.  As for $E_6$, this is too complex to perform
by inspection, so we work systematically through weight and
root lattices.

Following the conventions and notation of \cite{Chen:2018wep},
a basis for the weight lattice can be given as
\begin{eqnarray}
Y_1 & \sim & (0, 0, 0, 0, 0, 1, 0),
\\
Y_2 & \sim & (0, 0, 0, 0, 1, -1, 0),
\\
Y_3 & \sim & (0, 0, 0, 1, -1, 0, 0),
\\
Y_4 & \sim & (0, 0, 1, -1, 0, 0, 0),
\\
Y_6 & \sim & (0, 1, 0, 0, 0, 0, -1),
\\
Y_7 & \sim & (1, -1, 0, 0, 0, 0, 1),
\\
Y_{36} & \sim & (1, 0, 0, 0, 0, 0, -1).
\end{eqnarray}
We can then write a basis of root lattice vectors as
\begin{eqnarray}
X_1 & \sim & (1, 0, 0, 0, 0, 0, 0) \: = \: Y_6 + Y_7,
\\
X_2 & \sim & (-1, 1, 0, 0, 0, 0, 0) \: = \: Y_6 - Y_{36},
\\
X_3 & \sim & (0, -1, 1, 0, 0, 0, 0) \: = \: Y_1 + Y_2 + Y_3 + Y_4 - 2 Y_6 - Y_7
+ Y_{36},
\\
X_4 & \sim & (0, 0, -1, 1, 0, 0, 1) \: = \: - Y_4 + Y_6 + Y_7 - Y_{36},
\\
X_5 & \sim & (0, 0, 0, -1, 1, 0, 1) \: = \: - Y_3 + Y_6 + Y_7 - Y_{36},
\\
X_6 & \sim & (0, 0, 0, 1, 0, 0, -1) \: = \: 
Y_1 + Y_2 + Y_3 - Y_6 - Y_7 + Y_{36},
\\
X_7 & \sim & (0, 0, 0, 0, -1, 1, 1) \: = \:
- Y_2 + Y_6 + Y_7 - Y_{36},
\end{eqnarray}
which can be encoded in the matrix
\begin{equation}
A \: = \: \left[ \begin{array}{rrrrrrr}
0 & 0 & 0 & 0 & 1 & 1 & 0 \\
0 & 0 & 0 & 0 & 1 & 0 & -1 \\
1 & 1 & 1 & 1 & -2 & -1 & 1 \\
0 & 0 & 0 & -1 & 1 & 1 & -1 \\
0 & 0 & -1 & 0 & 1 & 1 & -1 \\
1 & 1 & 1 & 0 & -1 & -1 & 1 \\
0 & -1 & 0 & 0 & 1 & 1 & -1
\end{array} \right].
\end{equation}
A Smith decomposition of $A$ is given by
\begin{equation}
Q^{-1} \: = \: \left[ \begin{array}{rrrrrrr}
0 & -1 & -1 & -1 & 1 & 2 & 1 \\
1 & 0 & 0 & 0 & 0 & 0 & -1 \\
1 & 0 & 0 & 0 & -1 & 0 & 0 \\
1 & 0 & 0 & -1 & 0 & 0 & 0 \\
1 & 0 & -1 & -1 & 0 & 1 & 0 \\
1 & -1 & 0 & 0 & 0 & 0 & 0 \\
1 & -1 & -1 & -1 & 0 & 1 & 0
\end{array} \right], 
\: \: \:
P \: = \: \left[ \begin{array}{rrrrrrr}
1 & 0 & 0 & 0 & 0 & 0 & -1 \\
0 & 1 & 0 & 0 & 0 & 0 & -1 \\
0 & 0 & 1 & 0 & 0 & 0 & -1 \\
0 & 0 & 0 & 1 & 0 & 0 & -1 \\
0 & 0 & 0 & 0 & 1 & 0 & -1 \\
0 & 0 & 0 & 0 & 0 & 1 & -1 \\
0 & 0 & 0 & 0 & 0 & 0 & 1
\end{array} \right].
\end{equation}
It is straightforward to check that
\begin{equation}
\det Q^{-1} \: = \: 1 \: = \: \det P,
\end{equation}
and
\begin{equation}
Q^{-1} A P \: = \: {\rm diag}(1, 1, 1, 1, 1, 1, 2),
\end{equation}
as expected since the center of $E_7$ is ${\mathbb Z}_2$.

The invariant, the map to ${\mathbb Z}_2$, is given from the right column
of $P$ by
\begin{equation}
\gamma \cdot \vec{m} \mod 2,
\end{equation}
where in a basis of $Y$'s,
\begin{equation}
\vec{\gamma} \: = \: (-1, -1, -1, -1, -1, -1, +1).
\end{equation}
Converting to the basis in \cite{Chen:2018wep}, we have
\begin{equation}
\vec{\gamma} \: = \: (-2, -4, -4, -3, -2, -1, -3),
\end{equation}
which gives the invariant
\begin{equation}
\vec{\gamma} \cdot \vec{m} \: \equiv \: m_4 + m_6 + m_7 \mod 2.
\end{equation}
It is straightforward to check that for every root $\vec{m}$, in the basis used
in \cite{Chen:2018wep}, $\vec{\gamma} \cdot \vec{m} \equiv 0 \mod 2$.

As before, we take the $t_a$ to be of the form
\begin{equation}
t_a \: = \: \frac{2 \pi i}{2} m_a,
\end{equation}
where $m_a \in {\mathbb Z}$.  From the weight/root analysis above, we see
that the ${\mathbb Z}_2$ discrete theta angle defined by a set of $m_a$ is
\begin{equation}
m_4 + m_6 + m_7 \mod 2.
\end{equation}

We are now ready to determine for which discrete theta angle the
pure $E_7/{\mathbb Z}_2$ gauge theory has supersymmetric vacua.
Using the critical locus equations\footnote{
There are some typos in the factors multiplying $\sigma_7$ in
\cite[equ'n (6.16)]{Chen:2018wep}, which we have corrected when performing
the computation given here.
}
from \cite[section 7]{Chen:2018wep},
with exponents of $1/2$ as in equation~(\ref{eq:e7:crit}),
it is straightforward to compute that
\begin{equation}
\exp(-t_4 - t_6 - t_7) \: = \:
\frac{ R_4 R_7 R_{13} }{R_8 R_{12} R_{20} }
\frac{R_{18} R_{19}}{ R_{26}}
\frac{ R_{27} }{ R_{28} }
R_{29}
\frac{ R_{36}}{ R_{38}}
\frac{ R_{37}}{R_{40}}
\frac{R_{39}}{R_{48}}
\frac{R_{47}}{R_{50}}
\frac{R_{49}}{R_{51}}
\frac{R_{59}}{R_{58}}
\frac{R_{60}}{R_{61}}
R_{62},
\end{equation}
where each $R_n$ indicates a ratio of the form $X_n / X_{n+63}$,
and all such ratios are $-1$ on the critical locus,
from \cite[section 6]{Chen:2018wep}.
(We should emphasize that the fact that integer powers of the $R_n$ appear,
rather than square roots, is itself a highly nontrivial check, as the
critical locus equations themselves involve factors of $R^{1/2}$.)
Since there are $27$ such factors, an odd number of factors of $-1$, we see that
\begin{equation}
\exp(-t_4 - t_6 - t_7) \: = \: -1
\end{equation}
on the critical locus, so that supersymmetric vacua only exist in the
case
\begin{equation}
m_4 + m_6 + m_7 \: \equiv \: 1 \mod 2,
\end{equation}
i.e. for the nontrivial ${\mathbb Z}_2$ discrete theta angle.

So far we have determined for which discrete theta angle a pure
$E_7/{\mathbb Z}_2$ gauge theory has supersymmetric vacua.
If we repeat the same analysis of \cite[section 6.4]{Chen:2018wep},
one finds that the critical locus is described by seven twisted chiral
multiplets with vanishing superpotential, just as in the case of a
pure $E_7$ gauge theory.

Putting this together, and comparing to our earlier results,
a pure $E_7/{\mathbb Z}_2$ gauge theory has vacua (corresponding to seven
fields) in the case of a nontrivial
discrete theta angle.  We note this is consistent with decomposition:
a pure $E_7$ gauge theory decomposes as the sum of two pure
$E_7/{\mathbb Z}_2$ gauge theories with either value of the discrete
theta angle.  One has no supersymmetric vacua, the other evidently
flows to the same fixed point as the pure $E_7$ gauge theory, and so
decomposition is consistent.

\section{Summary and comparison with prior results}

Briefly, our results here together with the analysis of
\cite{Gu:2018fpm,Chen:2018wep} demonstrates that nonabelian mirrors
are consistent with the statement that pure (2,2) supersymmetric
$G$ gauge theories admit supersymmetric vacua for precisely one choice
of discrete theta angle, for which they flow in the IR to a set of
free twisted chiral superfields, as many as the rank of $G$,
and break supersymmetry in the IR for other values of the discrete
theta angle.
We will walk through the details, and also verify that the results are
consistent with decomposition.

First,
these results correctly reproduce the result of
\cite{Aharony:2016jki},
\cite[section 12.1]{Gu:2018fpm} 
that pure supersymmetric $SU(2)$ gauge theories and
$SO(3) = SU(2)/{\mathbb Z}_2$ theories with nontrivial discrete theta angle
have supersymmetric vacua, but the pure $SO(3)$ theory with vanishing
discrete theta angle does not have any supersymmetric vacua.

More generally, pure supersymmetric $SU(k)$ gauge theories have
\cite[section 12.3]{Gu:2018fpm} vacua, flowing in the IR to
$k-1$ degrees of freedom, as do $SU(k)/{\mathbb Z}_k$ gauge theories for
precisely one choice of discrete theta angle, namely
\begin{equation}
\sum_a m_a \: \equiv \: - \frac{k(k-1)}{2} \mod k,
\end{equation}
as we argued in section~\ref{sect:suk}.

Pure supersymmetric $SO(2k)$ gauge theories were argued 
in \cite[section 13.1]{Gu:2018fpm} to have supersymmetric vacua
(and $k$ IR degrees of freedom)
for exactly one choice of (${\mathbb Z}_2$)
discrete theta angle, namely the trivial one.
We studied pure $SO(2k)/{\mathbb Z}_2$ gauge theories in 
section~\ref{sect:so2k}, and argued that they also have supersymmetric vacua
(and $k$ degrees of freedom)
for exactly one choice of discrete theta angle (either ${\mathbb Z}_2 \times
{\mathbb Z}_2$ or ${\mathbb Z}_4$), given by
\begin{equation}
\sum_a m_a \: \equiv \: k(k-1) \mod 4, \: \: \:
m_1 \: \equiv \: 0 \mod 2.
\end{equation}

As a consistency check, note that
$SU(4)/{\mathbb Z}_4$ and  
$SO(6)/{\mathbb Z}_2$ both have supersymmetric
vacua (and the same number of IR degrees of freedom)
in the descriptions above for the same discrete theta angle,
\begin{equation}
\sum_a m_a \: \equiv \: 2 \mod 4,
\end{equation}
as expected since $SU(4)/{\mathbb Z}_4 = SO(6)/{\mathbb Z}_2$.
(Similarly, we checked in section~\ref{sect:su4:z2}) that
$SO(6)$ and $SU(4)/{\mathbb Z}_2$ also have supersymmetric vacua
for the same discrete theta angle, as expected since they are the 
same group.)

Pure supersymmetric $SO(2k+1)$ gauge theories were studied in
\cite[section 13.2]{Gu:2018fpm}, and argued to 
have supersymmetric vacua
for the one nontrivial discrete theta angle, with as many IR degrees of
freedom as the rank.
(Note that this is consistent with $SU(k)/{\mathbb Z}_k$ results,
which also flow in the IR to a free field theory for a nontrivial
discrete theta angle, as $SO(3) = SU(2)/{\mathbb Z}_2$.)
As $SO(2k+1)$ has no center, no further quotients exist.

Pure supersymmetric $Sp(2k)$ gauge theories were studied in
\cite[section 13.3]{Gu:2018fpm}, and argued to 
have supersymmetric vacua, with as many IR degrees of freedom as the rank.
We studied pure $Sp(2k)/{\mathbb Z}_2$ gauge theories
in section~\ref{sect:sp2k}, where we argued that they 
have supersymmetric vacua with as many degrees of freedom as the rank
for precisely one discrete theta angle.
This matches the results reviewed in section~\ref{sect:su2}
for pure $SO(3) = Sp(2)/{\mathbb Z}_2$ gauge theories, 
for which supersymmetric vacua exist for
the nontrivial discrete theta angle. 
This also matches the results in \cite[section 13.2]{Gu:2018fpm}
for pure $SO(5) = Sp(4)/{\mathbb Z}_2$ gauge theories,
for which supersymmetric vacua exist for the nontrivial discrete
theta angle.

Pure $E_6$ gauge theories were discussed in \cite[section 5.4]{Chen:2018wep},
which argued that they 
have supersymmetric vacua with six (twisted chiral) IR degrees of freedom.
We discussed pure $E_6/{\mathbb Z}_3$ gauge theories in
section~\ref{sect:e6}, where we discovered that these also flow to the
same IR theories for a single choice of discrete theta angle.

Similarly, pure $E_7$ gauge theories were discussed in
\cite[section 6.4]{Chen:2018wep}, which argued that they 
have supersymmetric vacua with seven (twisted chiral) degrees of freedom.
We discussed pure $E_7/{\mathbb Z}_2$ gauge
theories in section~\ref{sect:e7}, where we discovered that these also
flow to the same IR theories for a single choice of discrete theta angle.

Pure supersymmetric $G_2$, $F_4$, and $E_8$ gauge theories
were discussed in \cite{Chen:2018wep}, which argued that they 
have supersymmetric vacua and as many (twisted chiral) IR degrees of freedom
as the rank.
However, these groups have no center, so there
are no further quotients to consider.

These results, for $G$ and $G/K$ gauge theories, are intimately related:
a pure $G/K$ gauge theory flows in the IR to the same endpoint, the same
free field theory, as a pure $G$ gauge theory, but for a single value
of the $G/K$ discrete theta angle.  This is consistent with
nonabelian decomposition in two-dimensional theories
\cite{Hellerman:2006zs,Sharpe:2014tca,Sharpe:2019ddn}:  
briefly, a two-dimensional
$G$ gauge theory with center $K$ `decomposes' into a sum of
$G/K$ gauge theories with various discrete theta angles.
Schematically:
\begin{equation}
\mbox{$G$-gauge theory} \: = \: \oplus_{\alpha \in K} \left(
\mbox{$G/K$-gauge theory}\right)_{\alpha}.
\end{equation}
(Technically this decomposition reflects the existence of a $BK$ one-form
symmetry in the original two-dimensional theory;
see \cite{Tanizaki:2019rbk} 
for an analogous decomposition in four-dimensional theories
with three-form symmetries.)
Any IR endpoint of the $G$ gauge theory must therefore be shared amongst
the $G/K$ gauge theories appearing in the decomposition, which must
reproduce that endpoint amongst their sum.  The results above are
consistent with this decomposition:  for each $G$ above,
exactly one of the $G/K$ theories (corresponding to one discrete theta
angle) has an IR fixed point, a free field theory,
the same as that of the $G$ gauge
theory.  

We can also use decomposition to make predictions for the IR behavior
of pure Spin$(n)$ gauge theories, based on our results for 
pure $SO(n)$ theories.  Recall pure $SO(n)$ theories with
precisely one discrete theta angle
have an IR limit (a free theory, with as many
twisted chiral superfields as the rank), and the other discrete
theta angles prohibit any supersymmetric vacua.
Since a Spin$(n)$ theory decomposes as a union of those $SO(n)$ theories,
we see that a pure supersymmetric two-dimensional
Spin$(n)$ theory flows in the IR to the same free theory, with as
many twisted chiral superfields as the rank.

As consistency checks, let us compare some special cases:
\begin{itemize}
\item Spin$(3) = SU(2)$.  It was argued in \cite{Aharony:2016jki}
that pure $SU(2)$ theories flow in the IR to a theory with one
twisted chiral superfield, consistent with the claim above
for Spin$(3)$.
\item Spin$(4) = SU(2) \times SU(2)$.  Each of the pure $SU(2)$ theories
flows in the IR to a theory of one twisted chiral
superfield, hence the Spin$(4)$ theory must flow in the IR to a theory
of two such superfields, which is consistent with the claim above
for Spin$(4)$.
\item Spin$(5) = Sp(4)$.  As discussed in \cite[section 13.3]{Gu:2018fpm},
a pure supersymmetric $Sp(4)$ gauge theory in two dimensions flows in the
IR to a theory of two twisted chiral multiplets,
which matches the prediction above for Spin$(5)$.
\item Spin$(6) = SU(4)$.  As discussed in \cite[section 12.3]{Gu:2018fpm},
a pure supersymmetric $SU(4)$ gauge theory in two dimensions flows in the 
IR to a theory of three twisted chiral multiplets,
which matches the prediction above for Spin$(6)$.
\end{itemize}

\section{Conclusions}

In this paper we have taken another step towards understanding
the IR behavior of pure two-dimensional (2,2) supersymmetric
gauge theories, following \cite{Aharony:2016jki,Gu:2018fpm,Chen:2018wep}.
Specifically, both directly in gauge theories and in parallel with
nonabelian mirror symmetry we have examined
pure $SU(k)/{\mathbb Z}_k$, $SO(2k)/{\mathbb Z}_2$, $Sp(2k)/{\mathbb Z}_2$,
$E_6/{\mathbb Z}_3$, and $E_7/{\mathbb Z}_2$ gauge theories.
The possible discrete theta angles
in these theories are considerably more complicated than those appearing
in examples studied previously 
in \cite{Aharony:2016jki,Gu:2018fpm,Chen:2018wep}, 
which we have worked out in detail, and in parallel we have also
extended and tested the technology of nonabelian mirrors.

\section{Acknowledgements}

We would like to thank M.~Ando, K.~Hori, and T.~Pantev for useful conversations.
E.S. was partially supported by NSF grant PHY-1720321.

\end{document}